\documentclass[twocolumn]{aastex631}

\usepackage{natbib}
\usepackage{graphicx}
\usepackage{hyperref}
\usepackage{apjfonts}
\usepackage{float}

\begin{document}

\newcommand{\LCO}{Las Cumbres Observatory, 6740 Cortona Dr Ste 102, Goleta, CA 93117-5575, USA}
\newcommand{\UCSB}{Department of Physics, University of California, Santa Barbara, CA 93106-9530, USA}
\newcommand{\KITP}{Kavli Institute for Theoretical Physics, University of California, Santa Barbara, CA 93106, USA}
\newcommand{\Rutgers}{Department of Physics and Astronomy, Rutgers, The State University of New Jersey, 136 Frelinghuysen Road, Piscataway, NJ 08854, USA}
\newcommand{\UCSC}{Department of Astronomy and Astrophysics, University of California, Santa Cruz, CA 92064, USA}
\newcommand{\CfA}{Center for Astrophysics \textbar{} Harvard \& Smithsonian, 60 Garden Street, Cambridge, MA 02138-1516, USA}
\newcommand{\Sydney}{School of Mathematics and Statistics, The University of Sydney, Camperdown, NSW 2006 Australia}
\newcommand{\CTDS}{Centre for Translational Data Science, University of Sydney, Darlington, NSW 2008, Australia}
\newcommand{\UT}{University of Texas at Austin, 1 University Station C1400, Austin, TX, 78712-0259, USA}
\newcommand{\STSci}{Space Telescope Science Institute, 3700 San Martin Drive, Baltimore, MD 21218, USA}
\newcommand{\JHU}{Department of Physics and Astronomy, Johns Hopkins University, Baltimore, MD 21218, USA}
\newcommand{\Moore}{Gordon and Betty Moore Foundation, 1661 Page Mill Road, Palo Alto, CA 94304}
\newcommand{\Yunnan}{Yunnan Observatories, Chinese Academy of Sciences (CAS), Kunming 650216, P.R. China}
\newcommand{\KeyCAS}{Key Laboratory for the Structure and Evolution of Celestial Objects, CAS, Kunming 650216, P.R. China}
\newcommand{\UCAS}{University of Chinese Academy of Science, Beijing 100012, P.R. China}
\newcommand{\MegaScience}{Center for Astronomical Mega-Science, CAS, Beijing 100012, P.R. China}

\title{Still Brighter than Pre-Explosion, SN 2012Z Did Not Disappear:\\ Comparing Hubble Space Telescope Observations a Decade Apart}

\author[0000-0001-5807-7893]{Curtis~McCully}
\altaffiliation{\href{mailto:cmccully@lco.global}{cmccully@lco.global}}
\affiliation{\LCO}
\affiliation{\UCSB}

\author[0000-0001-8738-6011]{Saurabh~W.~Jha}
\affiliation{\Rutgers}

\author{Richard~A.~Scalzo}
\affiliation{\Sydney}
\affiliation{\CTDS}

\author[0000-0003-4253-656X]{D.~Andrew~Howell}
\affiliation{\LCO}
\affiliation{\UCSB}

\author{Ryan~J.~Foley}
\affiliation{\UCSC}

\author{Yaotian~Zeng}
\affiliation{\Yunnan}
\affiliation{\KeyCAS}
\affiliation{\UCAS}

\author{Zheng-Wei~Liu}
\affiliation{\Yunnan}
\affiliation{\KeyCAS}
\affiliation{\MegaScience}

\author[0000-0002-0832-2974]{Griffin~Hosseinzadeh}
\affiliation{\CfA}

\author{Lars~Bildsten}
\affiliation{\UCSB}
\affiliation{\KITP}

\author{Adam~G.~Riess}
\affiliation{\STSci}
\affiliation{\JHU}

\author[0000-0002-1966-3942]{Robert~P.~Kirshner}
\affiliation{\CfA}
\affiliation{\Moore}

\author{G.~H.~Marion}
\affiliation{\UT}

\author{Yssavo~Camacho-Neves}
\affiliation{\Rutgers}

\begin{abstract}
Type Iax supernovae represent the largest class of peculiar white-dwarf supernovae. The type Iax SN~2012Z in NGC 1309 is the only white dwarf supernova with a detected progenitor system in pre-explosion observations. Deep \textit{Hubble Space Telescope} images taken before SN~2012Z show a luminous, blue source that we have interpreted as a helium-star companion (donor) to the exploding white dwarf. We present here late-time \textit{HST} observations taken $\sim$1400 days after the explosion to test this model. We find the SN light curve can empirically be fit by an exponential decay model in magnitude units. The fitted asymptotic brightness is within $10\%$ of our latest measurements and approximately twice the brightness of the pre-explosion source. The decline of the light curve is too slow to be powered by $^{56}$Co or $^{57}$Co decay: if radioactive decay is the dominate power source, it must be from longer half-life species like $^{55}$Fe. Interaction with circumstellar material may contribute to the light curve, as may shock heating of the companion star. Companion-star models underpredict the observed flux in the optical, producing most of their flux in the UV at these epochs. A radioactively-heated bound remnant, left after only a partial disruption of the white dwarf, is also capable of producing the observed excess late-time flux. Our analysis suggests that the total ejecta + remnant mass is consistent with the Chandrasekhar mass for a range of type Iax supernovae.
\end{abstract}

\section{Introduction}\label{sec:introduction}
A thermonuclear supernova explosion of a white dwarf is supposed to be the terminal event in the life of its progenitor star. For a normal type Ia supernova (SN Ia), i.e., the kind used as a standardizable candle for cosmology, the explosion is thought to completely unbind the star. Nevertheless, it is now clear that there is a wider variety of white dwarf supernovae than just typical SN Ia \citep[e.g.,][]{Taubenberger17,Jha19}.

Type Iax supernovae (SNe Iax; \citealt{Foley13}) comprise the most populated class of peculiar white dwarf supernovae. Based on the prototype SN~2002cx \citep{Li03}, these are subluminous, low-velocity explosions (see \citealt{Jha17} for a review) compared to normal SNe Ia. In particular, there is mounting evidence that SNe~Iax may not fully destroy the star, but instead may leave behind a bound remnant, in contrast to the complete disruption expected in SNe Ia.

At late times, observations of SNe~Iax imply high density material that is not seen in normal SNe~Ia. About a year past peak, the light curves of SNe~Iax like SN~2005hk, SN~2008A, and SN~2014dt decline more slowly than those of normal SNe Ia, suggesting efficient $\gamma$-ray trapping from high density material \citep{McCully14_08A, Kawabata18}. At similar epochs, the spectra of SNe Iax still have P-Cygni features, requiring an optically thick photosphere, contrasted with the fully nebular spectra of SNe~Ia \citep{Jha06, Sahu08, McCully14_08A, Stritzinger15}.

SN~2005hk showed extremely narrow forbidden lines of just 500 $km / s$ \citep{McCully14_08A} suggesting that either the explosion was finely tuned to just barely unbind the star or that the explosion did not fully disrupt the progentior. \citet{Foley16} suggest a two-component model to explain the late-time spectra of SNe Iax. The broader, forbidden lines are produced by the SN ejecta, while the P-Cygni lines are from a wind driven by the remnant star, powered by the radioactive decay of $^{56}$Ni produced in the SN explosion.  Given this hypothesis, \citet{Shen17} model the light-curve of such a remnant and find that the flux of the remnant should dominate that of the SN ejecta a few years after explosion. \citet{Vennes17} and \citet{Raddi19} have argued that they have even observed such a remnant. We have also previously observed a coincident, red source with SN 2008ha, an extreme SN Iax, $\sim$4 years after explosion which could be such a remnant \citep{Foley14}. How this remnant is produced is still an open question. Pure deflagration models, for example, may not produce the layered ejecta seen in SN~2012Z (\citealt{Stritzinger15, Barna18}, but see \citealt{Magee2022} for a discussion of the layered ejecta). The data presented here will provide further constraints on possible explosion models. 

The type Iax SN~2012Z is unique as the only known white dwarf supernova with a detection of its progenitor system in pre-explosion images \citep{McCully14_12Z}. We discovered a luminous ($M_V \approx -5.3$) and blue ($B-V \approx -0.1$) source coincident with the supernova position, and argued that this was a single degenerate system, with a helium star companion that supplied mass to the accreting white dwarf that exploded\footnote{Similar analysis of pre-explosion observations of SN~2014dt \citep[see][for a discussion of the SN properties]{Singh18} did not detect a progenitor candidate \citep{Foley15}, so SN 2012Z's progenitor system may have been especially luminous or the progenitor system of SN~2014dt may have been too blue to have been detected.} (at some phase we expect the shock heated companion star to contribute the majority of the observed flux in the system). This explanation is supported by evidence of helium in some SN~Iax systems (\citealt{Foley13,Jacobson-Galan19, Magee19}; but see \citealt{White15}) and their young inferred age \citep{Foley09,Valenti09,Perets10,Lyman13,Foley13,McCully14_12Z,Lyman18,Takaro20} expected in helium-star donor models \citep[e.g.,][]{Claeys14,Wang17}. SN~2014dt did reveal a mid-IR excess compared to normal SNe Ia which could be attributed to a dusty CSM or a bound remnant \citep{Fox16}.

To test both the progenitor system model and prediction of a bound remnant in SN~Iax, we now turn to very late photometric observations of SN~2012Z. In this work, we present \textit{HST} observations of SN~2012Z out to more than 1400 days past peak brightness and roughly 10 years after the pre-explosion images described above. At these latest epochs, the interpretation of the photometric measurements is difficult: there are several possible contributing components to the flux. The pre-explosion flux could be a composite of flux from the companion star, the WD (including possible surface burning), and the accretion disk. At late times, we expect the accretion flux and the surface burning of the WD to no longer be present. However, the companion may have brightened due to interaction with the ejecta \cite[e.g.,][]{Pan13}. Furthermore, there may also be emission from the SN ejecta and from a bound remnant. If there is circumstellar material (CSM) that was ejected during the evolution of the progenitor system, interaction will eventually dominate the light curve \citep{Gerardy04, Graham19}. We consider each of these possibilities in our analysis below.

Throughout this work, we adopt a redshift for the host-galaxy of SN~2012Z, NGC 1309, of $z = 0.007125$ \citep[via NED \footnote{The NASA/IPAC Extragalactic Database (NED) is operated by the Jet Propulsion Laboratory, California Institute of Technology, under contract with the National Aeronautics and Space Administration.}]{HIPASS}. We take the Cepheid distance to NGC 1309 from \citet{Riess11}, $d = 33.0 \pm 1.4$ Mpc, which corresponds to a distance modulus of $\mu = 32.59 \pm 0.09$ mag. Throughout our analysis we use the early-time ground-based light curve of SN~2012Z from \citet{Stritzinger15}, and adopt their estimates of the date of $B$-band maximum light ($55967.39 \pm 0.11$ MJD) and the date of explosion ($55952.8 \pm 1.5$ MJD).

\section{Observations}\label{sec:observations}

\begin{figure}[t]
\includegraphics[width=\columnwidth]{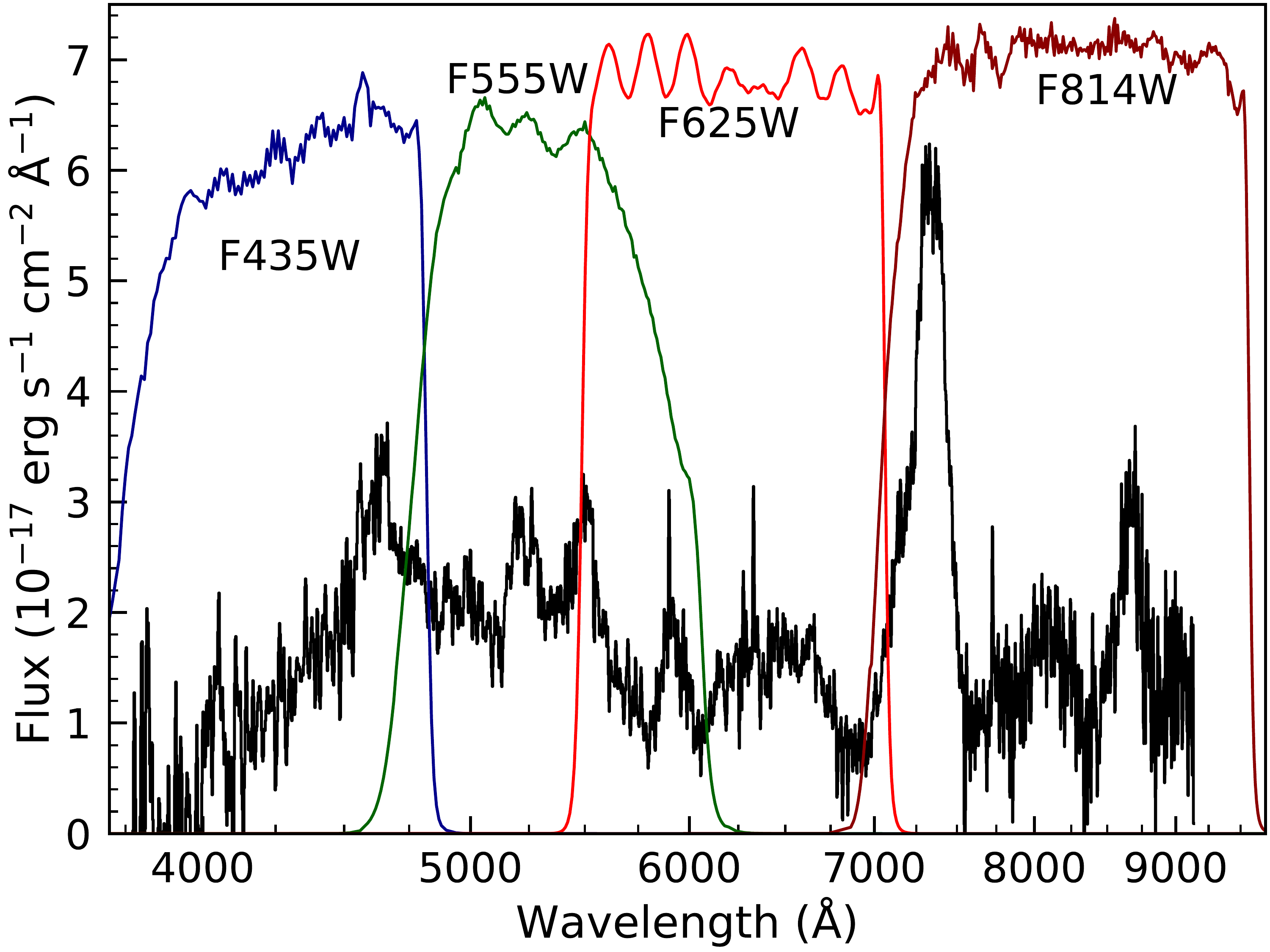}
\caption{Late-time spectra of SN 2012Z with the \textit{HST} bandpasses of interest. The spectrum was taken at $+215$ days past $B$-band maximum \citep{Stritzinger15}, but is nonetheless hundreds of days before the \textit{HST} photometry presented in this paper. The transmission curves of the following filters are shown: F435W, F555W, F625W, and F814W, which roughly correspond to $B$, $V$, $r$, and $I$, respectively. Both the \ion{Ca}{2} forbidden doublet and calcium-IR triplet drive the emission in F814W. If [\ion{O}{1}] $\lambda 6300$ was present (implying large amounts of unburned material), it would appear almost exclusively in the F625W filter. At this epoch, the emission is not well described by a blackbody. However, this spectrum was taken more than a thousand days before our latest photometry measurements. We note that SN 2012Z has higher velocities and does not resolve the large number of Fe P-Cygni lines seen in SN 2005hk \citep{Foley16,McCully14_08A}.\label{fig:bandpasses}}
\end{figure}

Pre-explosion images of NGC 1309 were taken under HST-GO-10497 (PI: Riess), HST-GO-10802 (PI: Riess), HST-GO-11570 (PI: Riess), and HST-GO-10711 (PI: Noll). Our late time observations of SN 2012Z were taken with \textit{HST} under the programs HST-GO-12913, HST-GO-13360, and HST-GO-13757 (PI: Jha). Images were obtained using the WFC3/UVIS and ACS/WFC instruments. Figure \ref{fig:bandpasses} shows one of the latest spectra of SN 2012Z (+215 days; \citealt{Stritzinger15}) and the \textit{HST} bandpasses that were used: F435W, F555W, F625W, and F814W (roughly, $B$, $V$, $r$, and $I$). Note that this spectrum was taken $>1000$ days before our latest HST observations.

We use the ``FLC'' frames that have had the pixel-based charge transfer efficiency (CTE) correction \citep{Anderson10} applied throughout our analysis.

The key step of our reduction is the astrometric registration of the \textit{HST} images, as they are taken across different cameras and across several visits. Each individual visit was astrometrically aligned to the pre-explosion \textit{HST} images used in \citet{McCully14_12Z}. The registration was done in two passes. The coarse registration was done by hand by measuring aperture photometry of overlapping stars and using the \textit{geomap} and \textit{geotrans} routines in IRAF. The fine registration was then done using the TweakReg task in DrizzlePac \citep{Drizzlepac}. To improve the registration, we create a set of temporary images that have had cosmic rays removed using Astro-SCRAPPY (\citealt{Astroscrappy}; see also \citealt{vanDokkum01}).

\begin{figure*}[th]
\begin{center}
\includegraphics[width=\textwidth]{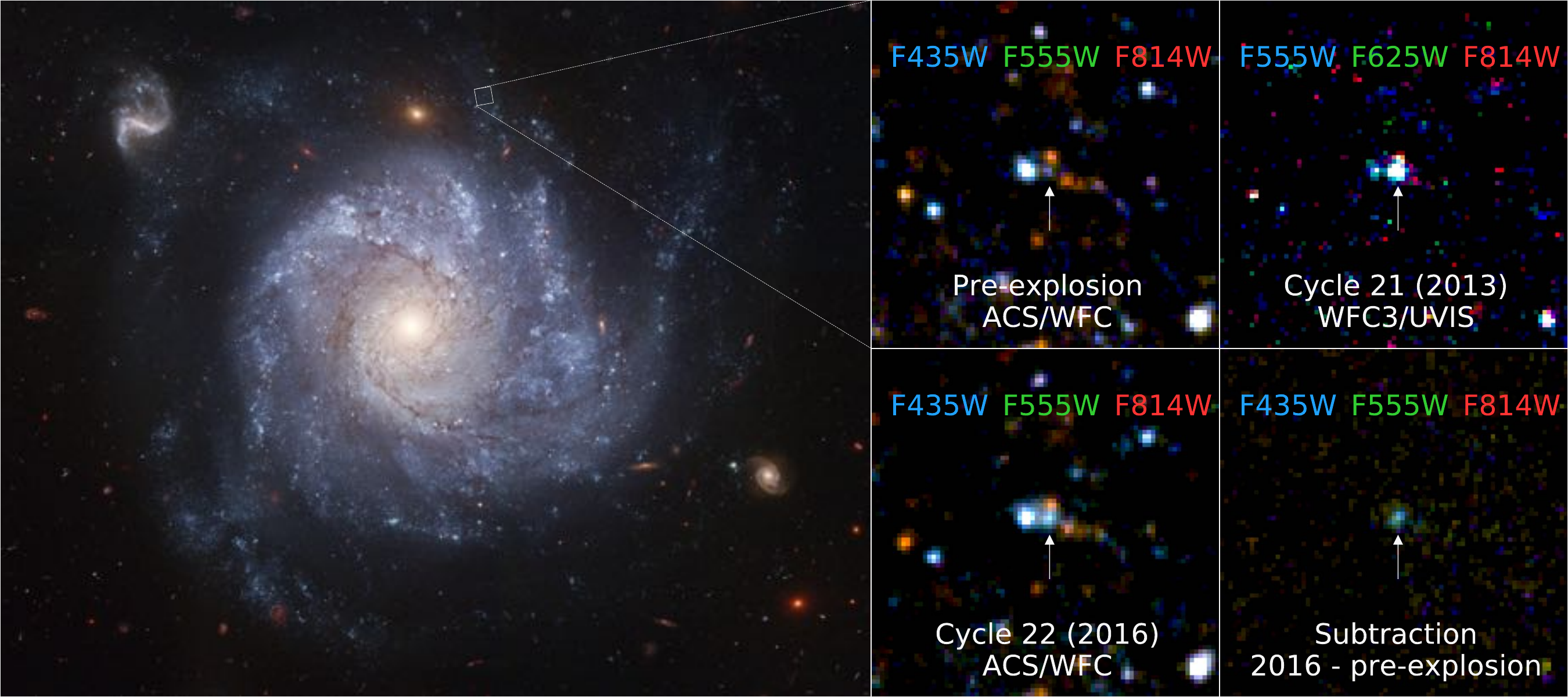}
\end{center}
\caption{Color images of NGC 1309 both before and after SN~2012Z. The left panel shows the Hubble Heritage (pre-explosion) image of NGC 1309  (\url{http://heritage.stsci.edu/2006/07}). The top middle panel shows a zoom in on the position of the supernova from the pre-explosion image. The top right shows SN~2012Z from the 2013 visit. The middle bottom panel shows the location of SN~2012Z in our latest observations in 2016. The bottom right panel shows the difference image between the pre-explosion images and the observations from 2016. The source at the location of SN~2012Z is still significantly brighter in 2016 than in the pre-explosion images. The difference is largest in the F555W filter, giving the source in the subtraction image a green hue.}
\label{fig:colorimage}
\end{figure*}

The original (non-cosmic ray rejected) images were resampled onto a common pixel grid and combined, with cosmic ray rejection, using Astrodrizzle also provided by DrizzlePac \citep{Drizzlepac}. The combined images are shown in Figure \ref{fig:colorimage}. The left panel shows the Hubble Heritage image of NGC 1309 (\url{http://heritage.stsci.edu/2006/07}). The top middle panel shows the pre-explosion image from \citet{McCully14_12Z}. The top right panel shows SN~2012Z in 2014. The bottom middle panel shows the latest observation of NGC 1309 from 2016 with the difference between the latest observation and the pre-explosion images shown in the bottom right panel. The source is clearly detected in all images and is still significantly brighter in the latest observations than in the pre-explosion image.

Photometry of the supernova and the field stars was performed using Dolphot \citep{Dolphot}, the updated version of HSTPhot \citep{Dolphin00}. Dolphot operates directly on the flat-fielded images (we use the flat-fielded and charge transfer efficiency corrected FLC files). Dolphot was run simultaneously on all \textit{HST} images that were taken in the same visit. \textit{HST} photometry of SN~2012Z and the pre-explosion source are given in Table \ref{tab:hstphot}.

\begin{deluxetable*}{ccccccccc}[ht]
\tablecaption{\textit{HST} Photometry of Pre-explosion Source S1 and SN~2012Z \label{tab:hstphot}}
\tablehead{\colhead{UT Date} & \colhead{Phase} & \colhead{Exposure Start} & \colhead{Exposure} & \colhead{Exposure} & \colhead{Instrument} & \colhead{Detector} & \colhead{Filter} & \colhead{Magnitude}\\ 
\colhead{} & \colhead{(days)} & \colhead{(MJD)} & \colhead{End (MJD)} & \colhead{Time (s)} & \colhead{} & \colhead{} &
\colhead{} & \colhead{(mag)}
} 
\startdata
2005-09-19 & $-2319$ & 53631.74 & 53632.84 & 9600  & ACS & WFC & F435W & $27.589 \pm 0.122$ \\
2006-03-16 & $-2142$ & 53588.65 & 54032.61 & 61760 & ACS & WFC & F555W & $27.622 \pm 0.060 $\\
2005-09-01 & $-2336$ & 53588.79 & 53640.63 & 24000 & ACS & WFC & F814W & $27.532 \pm 0.135$ \\
2010-07-24 & $-562$  & 55401.02 & 55401.23 & 6991 & WFC3 & IR & F160W & $26.443 \pm 0.321$ \\
2013-06-30 & $+502$ & 56473.96 & 56473.97 & 700 & WFC3 & UVIS & F555W & $24.369 \pm  0.034$ \\
2013-06-30 & $+502$ & 56473.95 & 56473.98 & 562 & WFC3 & UVIS & F625W & $24.333 \pm  0.050$ \\
2013-06-30 & $+502$ & 56473.95 & 56473.97 & 700 & WFC3 & UVIS & F814W & $24.469 \pm  0.079$ \\
2014-06-30 & $+865$ & 56838.85 & 56838.99 & 3002 & WFC3 & UVIS & F555W & $26.015 \pm  0.047$ \\
2014-06-30 & $+865$ & 56838.84 & 56838.98 & 1952 & WFC3 & UVIS & F625W & $26.009 \pm  0.079$ \\
2014-06-30 & $+865$ & 56838.86 & 56839.00 & 2402 & WFC3 & UVIS & F814W & $26.366 \pm  0.183$ \\
2016-01-16 & $+1425$ & 57392.46 & 57414.29 & 9624 & ACS & WFC & F435W & $27.059 \pm  0.058$ \\
2016-01-16 & $+1425$ & 57392.59 & 57414.42 & 12642 & ACS & WFC & F555W & $26.672 \pm  0.040$ \\
2016-01-16 & $+1425$ & 57392.53 & 57414.36 & 12868 & ACS & WFC & F814W & $26.682 \pm  0.063$ \\
\enddata
\end{deluxetable*}

Throughout our analysis, we compare to SN~2005hk and SN~2008A, two SNe Iax with very late \textit{HST} observations, SN~2008ha, a peculiar SN~Iax and the normal type~Ia SN~2011fe. The data for SN~2005hk come from \citet{Phillips07}, \citet{Sahu08}, \citet{Hicken09}, and \citet{McCully14_08A}. The data for SN~2008A is from \citet{Ganeshalingam11}, \citet{Hicken12}, and \citet{McCully14_08A}. We use the data for SN~2011fe from \citet{Munari12}, \citet{Richmond12}, \citet{Tsvetkov13}, \citet{Zhang16}, \citet{Shappee17}, and \citet{Kerzendorf17}. The early data from SN~2008ha shown here are from \citet{Valenti09} and \citet{Foley09}. The latest measurements are from HST and are presented in \citet{Foley14}.

\section{Analysis}\label{sec:analysis}
\subsection{Late-time Light Curve Behavior}
\begin{figure*}[t]
\includegraphics[width=\textwidth]{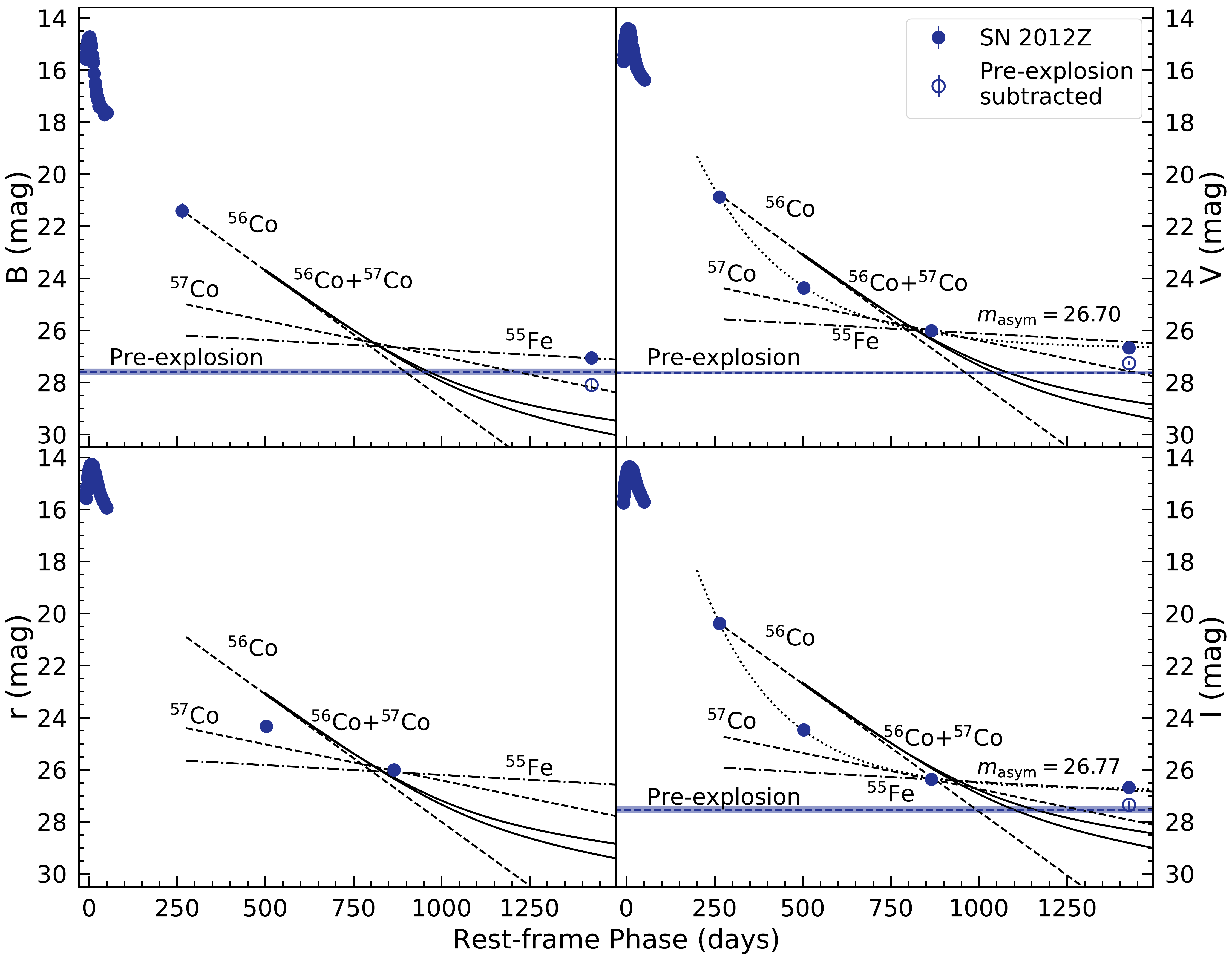}
\caption{\textit{BVri} (top left, top right, bottom left, and bottom right, respectively) light curves of SN~2012Z. The blue bands show the photometry of the progenitor system in the F435W, F555W, and F814W bands respectively. No filter corrections have been applied here. The latest three sets of points correspond to our \textit{HST} observations: F435W, F555W, F625W, and F814W roughly correspond to \textit{BVRI}, respectively. The unfilled (filled) circles show the photometry with (without) subtracting the pre-explosion flux. The dashed lines show the radioactive decay of $^{56}$Co and $^{57}$Co. The solid curves show light curves from explosion models from \citet{Ropke12}. In these models, the light curve is powered at late times by $^{57}$Co and then at the latest times by $^{55}$Fe. We have scaled the radioactive decay lines to match the observation in $V$ at $\sim 300$ days past explosion.  In all filters, the photometry from the latest \textit{HST} observations is several magnitudes above the $^{56}$Co line and a couple of magnitudes above the light curves from simulations. The points are also about a magnitude brighter than the photometry from the pre-explosion images. For $V$ and $I$, for which we have the most complete light curves, we find that in both cases the decline is too slow to be explained by even $^{57}$Co (disfavored by $3.8 \sigma$ and $3.5\sigma$ for $V$ and $I$ respectively or $5.2 \sigma$ jointly) so if the light curves are powered by radioactivity it must be from an element with a longer half-life like $^{55}$Fe. Empirically, the $V$ and $I$ light curves are well fit by an exponential decay model (in magnitude units) treating the decline time scale as a free parameter. We find an asymptotic magnitude of 26.70 and 26.77 respectively, within $2-\sigma$ of our latest measurements. We note that future observations may not follow this trend as new power sources become important.}
\label{fig:lightcurve}
\end{figure*}

Our late-time \textit{HST} observations of SN~2012Z yielded several surprising results. In Figure \ref{fig:lightcurve} we show the light curve of SN~2012Z for nearly 1500 days. We show the early data from \citet{Stritzinger15} taken in the \textit{BVri} filters. For the last three epochs, we show our \textit{HST} observations. The filters F435W, F555W, F625W, and F814W are similar to the ground based \textit{BVri} filters, respectively, so we do not apply a filter correction. Filter corrections require some estimate of the spectral energy distribution (SED) and are generally smaller ($\sim$0.1 mag) than the effects we are interested in here. 

The light curves for $V$ and $I$ are the most complete so we focus on those. The final four epochs for both filters are fit well to within the errors by an exponential decay model (in magnitude units) with an asymptotic magnitude of 26.70 and 26.77 for $V$ and $I$ respectively. These fits are shown as dotted lines in Figure \ref{fig:lightcurve}. For both filters, our latest observation is within $2-\sigma$ of the asymptotic magnitudes. This does not imply that this trend will continue at later epochs when new energy sources may become important. 

The blue bands show the photometry from the pre-explosion images from \citet{McCully14_12Z}. The final epoch of photometry is taken with the same filters and instrument as the pre-explosion images, so these can be directly compared without any filter correction. We find that in all bands, even at nearly 1500 days past explosion, SN~2012Z is still brighter than our photometry from pre-explosion images. We thus consider it extremely unlikely that the source seen in the pre-explosion images was the star that exploded as SN~2012Z as the source has not disappeared as has been observed for core-collapse SNe. Disentangling the contributions to the late-time flux and whether the progenitor system remained unchanged is more complicated.

\subsubsection{Radioactive Decay}
The black lines in Figure \ref{fig:lightcurve} show radioactive decay powered light curve models. The dashed line shows $^{56}$Co, which is the generally accepted power source for light of normal SNe Ia until about a year past maximum brightness \citep{Pankey62}. However, at the latest times, other radioactive isotopes begin to play an important role: first $^{57}$Co, then $^{55}$Fe at the latest phases. The solid black lines show the results of 3D explosion simulations from \citet{Ropke12}. The two solid lines correspond to two different explosion models: a delayed detonation of a Chandrasekhar white dwarf and the violent merger of two white dwarf stars.  We have scaled the radioactive decay models to the $V$-band point taken $\sim$300 days past explosion.

We find that in all bands, the measured photometry of SN~2012Z is several magnitudes brighter than the $^{56}$Co lines and a few magnitudes above the models from \citet{Ropke12} that include other radioactive isotopes. These models implicitly have assumptions about the ratio of $^{57}$Co to $^{56}$Co built in so we consider the extreme case if all of the luminosity is produced by $^{57}$Co. These are shown as dashed lines in Figure \ref{fig:lightcurve}. We use the second to last epoch to set the normalization in both the $V$ and $I$ bands for which we have the best coverage. In the case that we do not subtract the pre-explosion flux, the pure $^{57}$Co models are ruled out at high significance. If we subtract the pre-explosion flux, the $^{57}$Co-only models are disfavored by $3.8\sigma$ and $3.5\sigma$ for $V$ and $I$ respectively. Jointly, these values disfavor the $^{57}$Co-only model at $5.2\sigma$.  Therefore, if radioactivity in the ejecta is powering the light curve at these latest phases, it must be dominated by elements with a longer half-life than $^{57}$Co like $^{55}$Fe. If we assume all of the luminosity in the final data point (see below for a discussion of integrating the bolometric luminosity), we find that $\approx0.1M_\odot$ of $^{55}Fe$ would be required to power the light curve. As stable Ni is observed in late time spectra \citep{Foley16}, one would expect some amount of $^{55}$Fe to be produced. We leave it to more physical models test if this much $^{55}$Fe can be produced along with the radioactive nickel masses and low ejecta masses we find below. 

\subsubsection{Comparison to other White-Dwarf SNe}
\begin{figure*}[t]
\begin{center}
\includegraphics[width=1.0\textwidth]{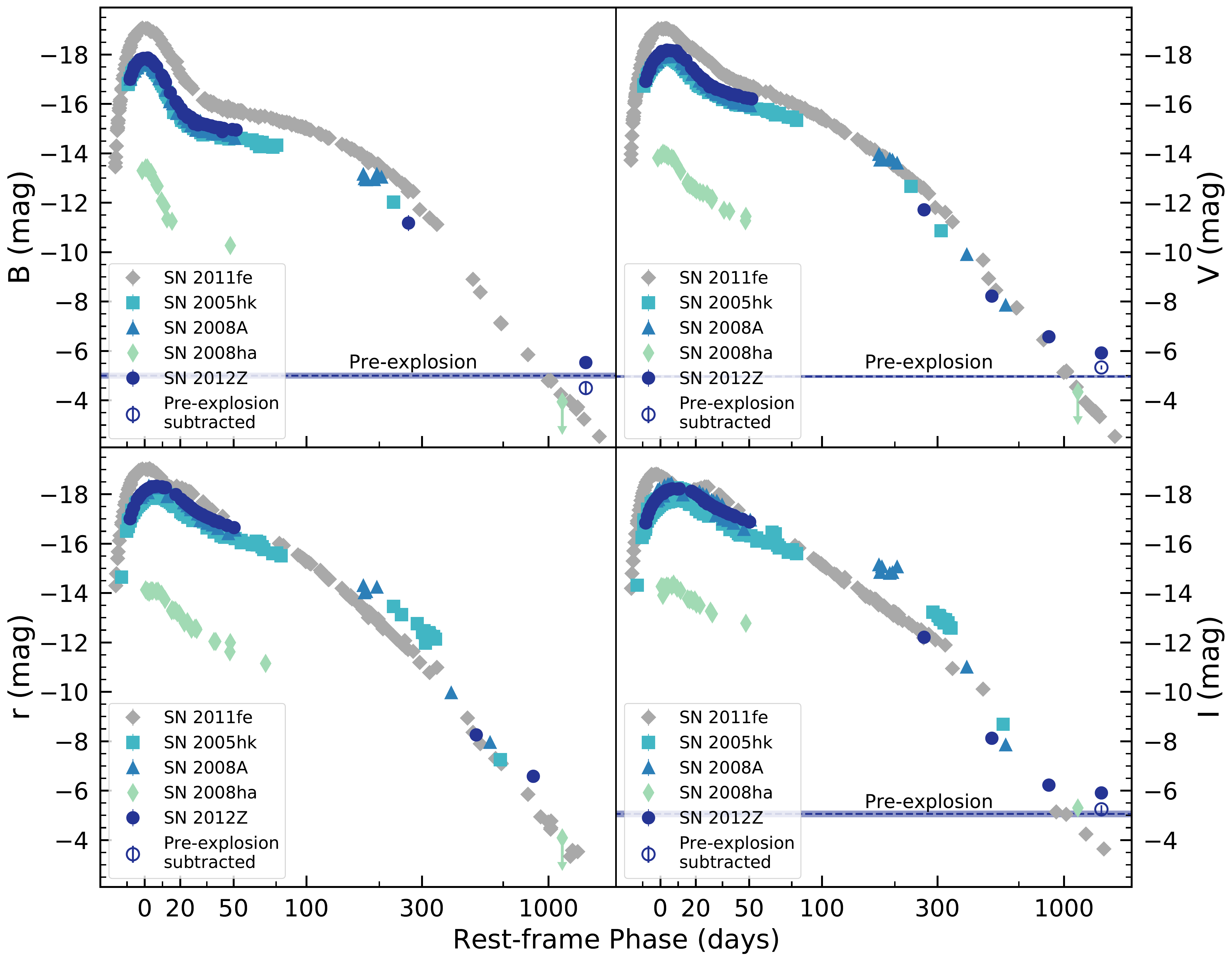}
\end{center}
\caption{Light curves of the type Iax SN~2012Z (dark blue circles), SN~2005hk (light blue squares), SN~2008A (light blue triangles), the extreme Iax SN~2008ha (green diamonds) and the normal Ia SN~2011fe (gray diamonds). We adopt this color/symbol scheme for the remainder of this work. The time axis employs an arcsinh scaling, which is approximately linear at early times but logarithmic at late times, to be able to compare the light curves of our sample of SNe at all phases. This scaling is used for the remainder of this work. At early times, SN~2012Z is brighter than the other SNe Iax shown here, but about a magnitude fainter than SN~2011fe. Between a year and two years after peak, SN~2012Z is slightly fainter than the other SNe Iax at similar epochs, but is comparable to SN~2011fe. At the latest phases, SN~2012Z is $\sim 2$ mag brighter than SN~2011fe. The light curve of SN~2012Z flattens out while SN~2011fe continues to decline. The source in the latest observations of SN 2008ha is fainter and is significantly redder than the latest epochs of SN~2012Z.}
\label{fig:lightcurve_compare}
\end{figure*}

The light curves of SN~2012Z are compared to the Iax SN 2005hk and SN~2008A, the peculiar Iax SN 2008ha, and the normal Ia SN~2011fe in Figure \ref{fig:lightcurve_compare}. SN~2005hk and SN~2008A have observations out to $\sim 600$ days past maximum light. At similar epochs, SN~2012Z was fainter than both SN~2005hk and SN~2008A, which is surprising given that SN~2012Z was one of the brightest SNe Iax at maximum light.

SN~2011fe was $\sim 1$ mag brighter than SN~2012Z at peak, but around three years after peak, the light curves have crossed. At $\sim 1450$ days past peak, SN~2012Z is $\sim 2$ mag \emph{brighter} than SN~2011fe. It appears that the light curve of SN~2012Z is leveling off above the detection in the pre-explosion images, whereas SN~2011fe is continuing to decline, roughly following radioactive decay of $^{57}$Co and $^{55}$Fe \citep[see][]{Kerzendorf17}.

We compare to observations of SN~2008ha near 1500 days past maximum from \citet{Foley14}. The source we detected that is coincident with the position of SN~2008ha is fainter and substantially redder (it was only detected in the reddest bands; \citealt{Foley14}) than that for SN~2012Z. 

\subsection{Late-time Spectral Energy Distribution of SN~2012Z}

At these late phases, the interpretation of the photometry becomes more complicated as there can now be contribution from multiple sources including the SN ejecta, the companion star, interaction with CSM if present, and possibly a bound remnant with a wind. 

At these extremely late phases, spectroscopy is no longer possible. Instead, we rely on photometry and the evolution of the SED of SN~2012Z to constrain the physics of SNe Iax. We consider two main scenarios that should bracket the true behavior of the SN. The first assumes that the pre-explosion flux was dominated by accretion and therefore does not contribute to the late-time HST photometry. The other scenario assumes that the companion star was the main contributor to the pre-explosion flux and would therefore be unchanged after the SN explosion (or could even become brighter).

\begin{figure}
\begin{center}
\includegraphics[width=\columnwidth]{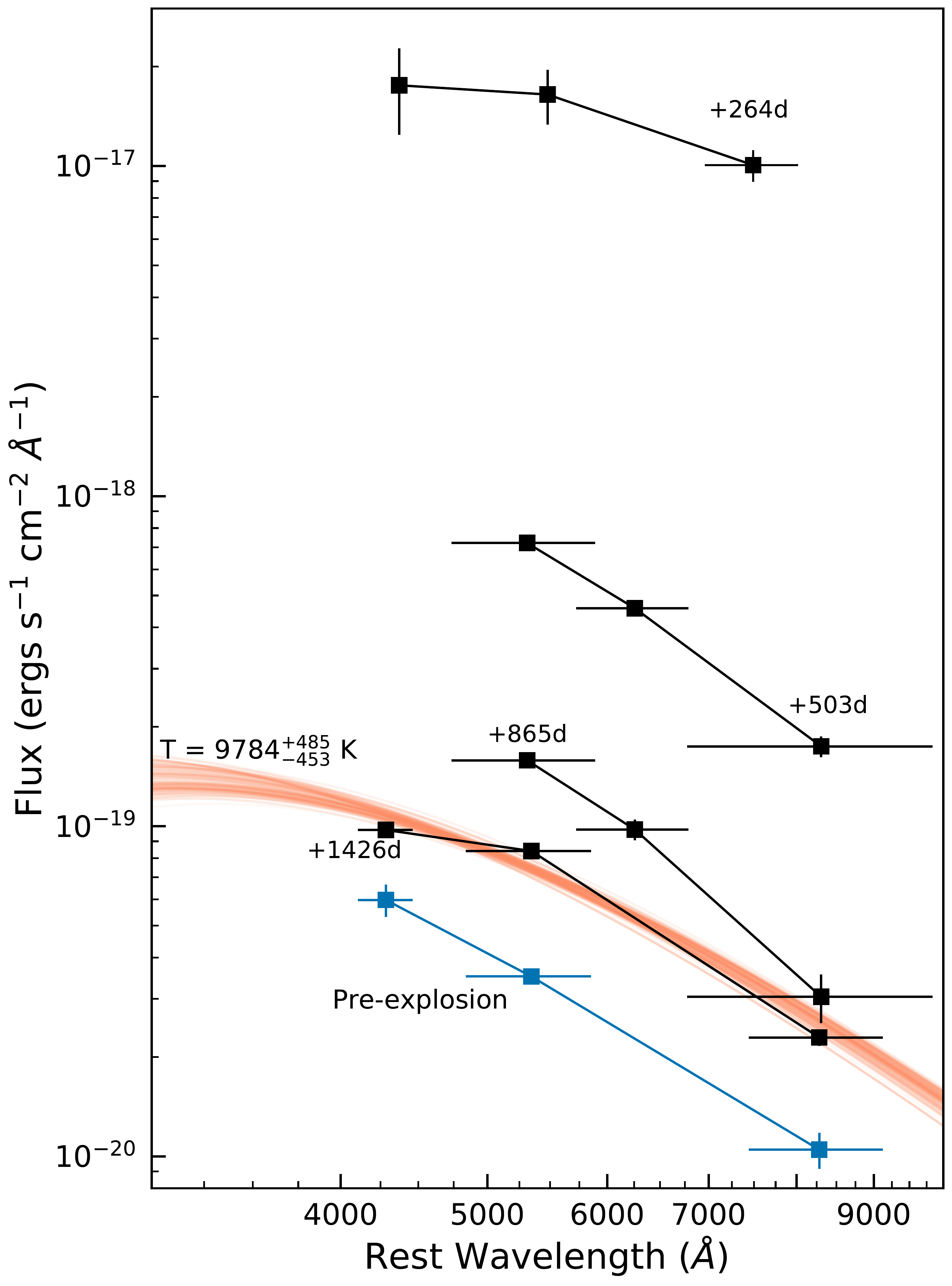}
\end{center}
\caption{\label{fig:sed} SED of the late-time \textit{HST} observations of SN 2012Z. The post-explosion points are shown in black while the pre-explosion points are shown in blue. The horizontal error bars show the width of the photometric passband. In the first \textit{HST} epoch at $\sim500$ days past maximum, the source is blue and consistent with being on the Rayleigh-Jeans tail with a temperature of $\gtrsim10,000$~K. At the latest epoch, more than 1400 days past maximum, the SN has become redder and is consistent with a blackbody temperature of $9784^{+485}_{-453}$K. The orange band shows the MCMC results for a blackbody fit to the latest epoch. This analysis assumes that the pre-explosion source is no longer present and does not contaminate the photometry at the latest epoch. We include the SED from $+264$ days when the spectrum is line dominated for comparison.}
\end{figure}

In Figure \ref{fig:sed}, we show the SED of SN 2012Z at the latest epochs taken with \textit{HST} without removing any contamination from the pre-explosion source. At $\sim 500$ days and $\sim 850$ days, the source is blue and has similar colors to the pre-explosion photometry. These measurements are consistent with being on the Rayleigh-Jeans tail of a blackbody, implying that we can only really constrain the temperature to be $\gtrsim 10,000$~K. The blue colors at late times suggest that SN 2012Z did not undergo an ``IR catastrophe'' with a redistribution of energy to the far infrared (\citealt{Sollerman04}, see discussion of the IR catastrophe in \citealt{McCully14_08A}).

If the late-time photometry has no contamination from the pre-explosion source, it would imply that the late-time emission is consistent with a blackbody. The latest spectrum we have of SN~2012Z (Figure \ref{fig:bandpasses}) is already non-thermal, so this could suggest that we are no longer measuring light from the SN ejecta, but are instead seeing some other component like the companion star or a bound remnant. 

\begin{figure}
\begin{center}
\includegraphics[width=\columnwidth]{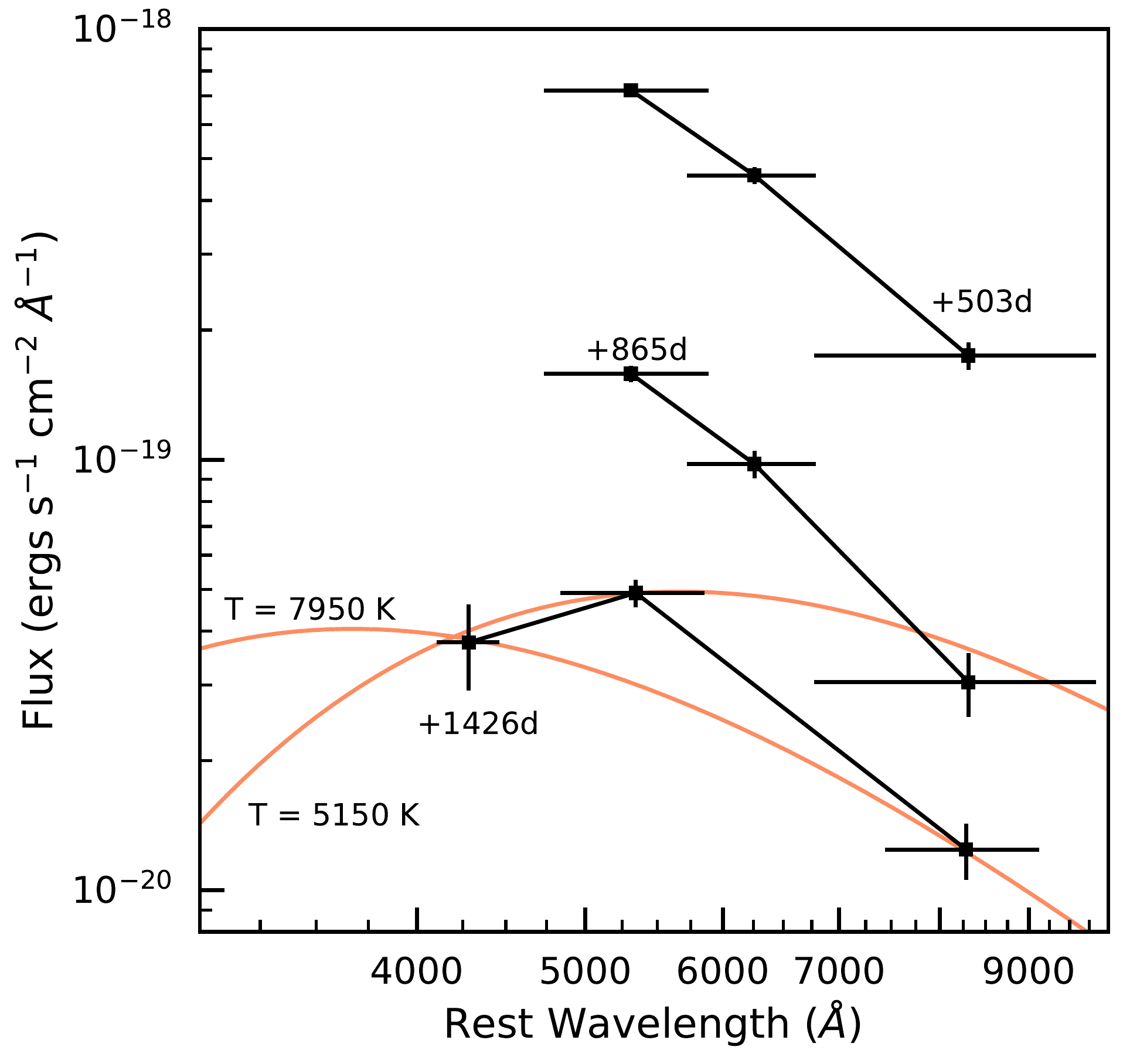}
\end{center}
\caption{\label{fig:sed_diff} SED of SN 2012Z after subtracting the flux from the pre-explosion source. Subtracting the pre-explosion flux has little effect on the earlier epochs because the SN/remnant is still bright relative to the pre-explosion detection. The final epoch is significantly different. The peak of the SED is consistent with a 5150~K blackbody, but this would predict more flux in the F814W band than we observe. The F435W and F814W points can be fit using a 7950~K blackbody, but this dramatically underpredicts the flux in F555W that we observe. If the pre-explosion source remains unchanged (e.~g.~the pre-explosion flux was dominated by the companion), then the SN/remnant is no longer dominated by thermal emission. The deviation from thermal is so extreme that it would require more than a solar mass of unburned He to explain the colors if we subtract the pre-explosion flux.}
\end{figure}

If we consider the case where the pre-explosion source is unchanged, we can take the difference of the fluxes from the pre-explosion images from our late-time observations of SN 2012Z. The SEDs are shown in Figure \ref{fig:sed_diff}. The earlier epochs (+503 and +865 days) are basically unchanged from above. The final epoch has a significantly different SED than the previous case. The SED is peaked in the F555W band, which corresponds to a blackbody temperature of 5000~K. While the F435W data is consistent with this temperature of blackbody, the F814W is more than a factor of two fainter than would be expected. This implies that if the pre-explosion source remained at constant brightness, then the radiation from the SN is non-thermal.

Based on the latest spectrum of SN~2012Z, we expect the spectrum from the SN ejecta to be line dominated which could explain the excess of V-band flux. Previously we searched for evidence of [O I] 6300\AA~\citep{McCully14_08A} from unburned oxygen predicted in pure deflagration models \citep{Kozma05}. We again see no evidence in any of our \textit{HST} photometry for an excess due to this line for SN~2012Z.

\citet{Foley13} suggest that SN~Iax could arise from He star donor systems, so we consider emission from He I 5876\AA~as a possible source of the excess V-band flux that could arise if He was stripped off of the donor star \citep[see][]{Zeng2020}. We fit a blackbody with a single He emission line to our photometry. We follow \citet{Jacobson-Galan19} using both their analytic model and results from \citet{Botyanszki18} to estimate the helium mass needed to explain our best-fit line flux. Both models require more than a solar mass of He making it unlikely that He I is the sole source of this flux excess.

The SED after subtracting the pre-explosion flux requires such strong line-flux that it is difficult to explain with any of the models we have tested. Given the unphysical amount of material needed to explain the excess V-band flux, we find that it is unlikely that the pre-explosion source is unchanged and therefore directly subtracting the pre-explosion flux is too simplistic. 

\subsection{Color Evolution of SN~2012Z}
We next compare the color evolution of SN~2012Z to other SNe Iax that have late-time observations and the canonical type~Ia SN~2011fe. One of the primary difficulties with comparing the light curves of these SNe is that the observations were taken in a range of filters. SN~2011fe was typically observed using Johnson filters while SN~2012Z was observed with \textit{BVgri} at early phases. While the \textit{HST} filters are approximately the same as the ground filters, there are small differences for which we need to account.

\begin{figure*}[t]
\begin{center}
\includegraphics[width=\textwidth]{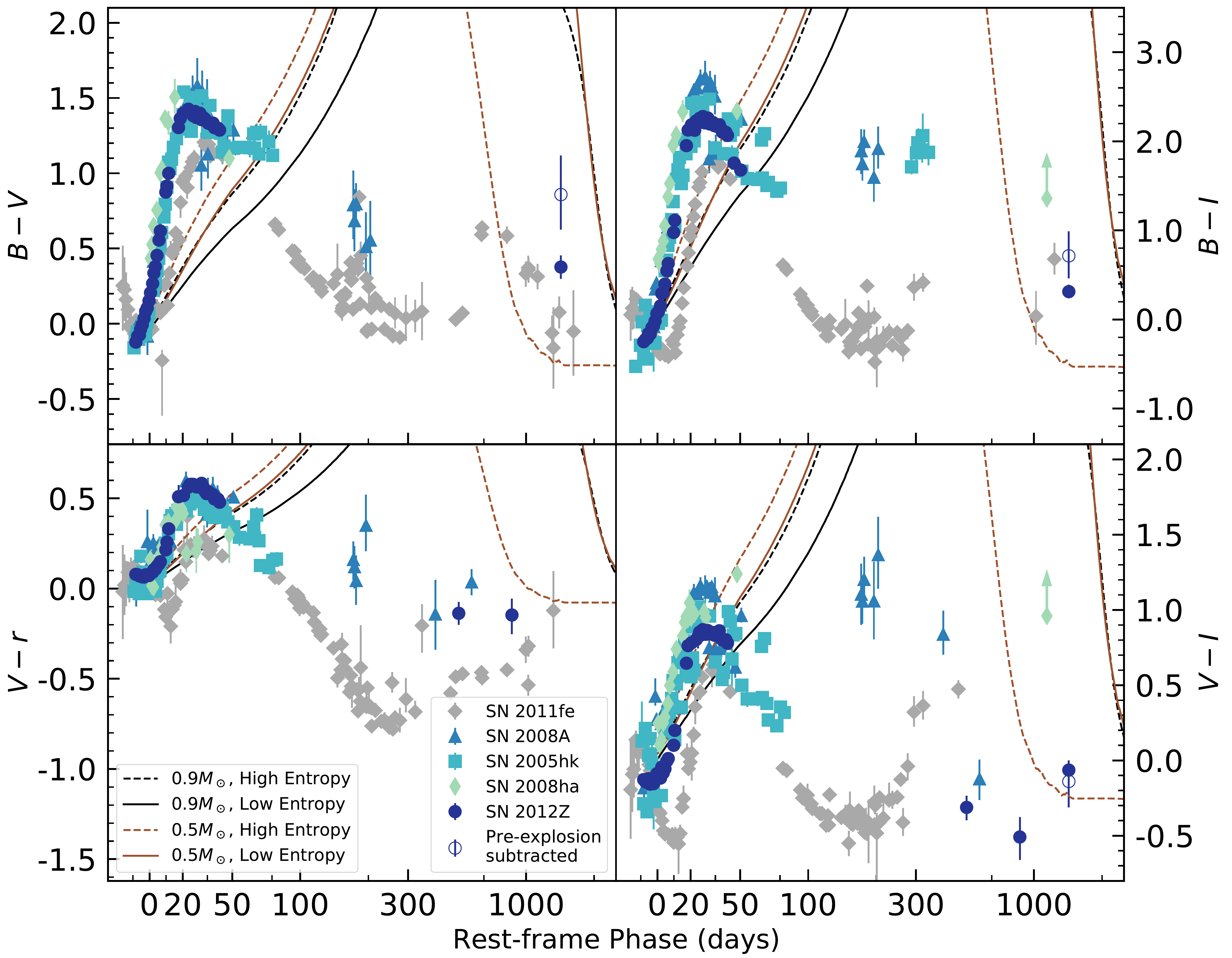}
\end{center}
\caption{Color evolution of the type Iax SN~2005hk, SN~2008A, SN~2008ha, and SN~2012Z, compared to the normal type~Ia SN~2011fe. Over the first $\sim 75$ days, the colors of the SNe Iax are similar to SN~2011fe. However, from $\sim 100$ days to $\sim 500$ days, the SNe Iax are redder than SN~2011fe. Then, SN~2012Z becomes bluer, reaching similar colors to SN~2011fe at the latest phases. Lines show the expected colors for models of a radioactively heated bound remnant from \citet{Shen17}. All of the models are too red to explain our latest observations except the high-entropy model of a 0.5 $M_\odot$ remnant. At early times, the colors are quite different than the models so are likely driven by a different mechanism, e.~g.~the SN ejecta.}
\label{fig:colorcurves_remnant}
\end{figure*}

To compare the colors across telescopes/filters, we model the SED of each SN at each epoch we have more than single band photometry. Our models are built by linearly interpolating between knots which are chosen to be at the center of the filter passbands. The fluxes of the knots are optimized to so that synthetic photometry matches the observed photometry. The uncertainties in the models are estimated using MCMC. We only consider colors for epochs that include observations in filters that have a central wavelength within 50 nm of the filters of interest. The color curves are shown in Figure \ref{fig:colorcurves_remnant}. We have included estimates of both pre-explosion subtracted and unsubtracted photometry (unfilled and filled markers respectively) for comparison. 

At early times, the colors of the SNe Iax are slightly redder but roughly follow the color evolution of SN~2011fe out to $\sim 75$ days. Around a hundred days past peak, normal SNe Ia become nebular, but SNe Iax do not \citep{Jha06, Foley13, McCully14_08A}. This key spectroscopic difference is also noticeable in the photometry. As SN~2011fe enters its nebular phase, it becomes bluer, while the SNe Iax remain red for several hundred days. 

In the latest \textit{HST} observations, SN~2012Z has become bluer again. Even by $\sim 500$ days past maximum it is much bluer than SN~2005hk and SN~2008A. The colors of SN~2012Z are consistent with the colors of SN~2011fe at the latest phases of $\sim 1500$ days past maximum. 

\citet{Foley16} suggest that a SN Iax that does not fully disrupt the white dwarf and could leave behind an observable remnant. \citet{Shen17} model the observational signatures of such a remnant. The newly produced radioactive material would heat the remnant, driving a wind. This could explain why SNe Iax show P-Cygni lines more than a year after explosion. We compare these models to our data in Figure \ref{fig:colorcurves_remnant}. Three of the four models from \citet{Shen17} are too red to be consistent with our observations. However, the high entropy model for a 0.5 $M_\odot$ remnant has similar colors to what we observe for SN 2012Z for the latest observation.

\begin{figure*}[!t]
\begin{center}
\includegraphics[width=\textwidth]{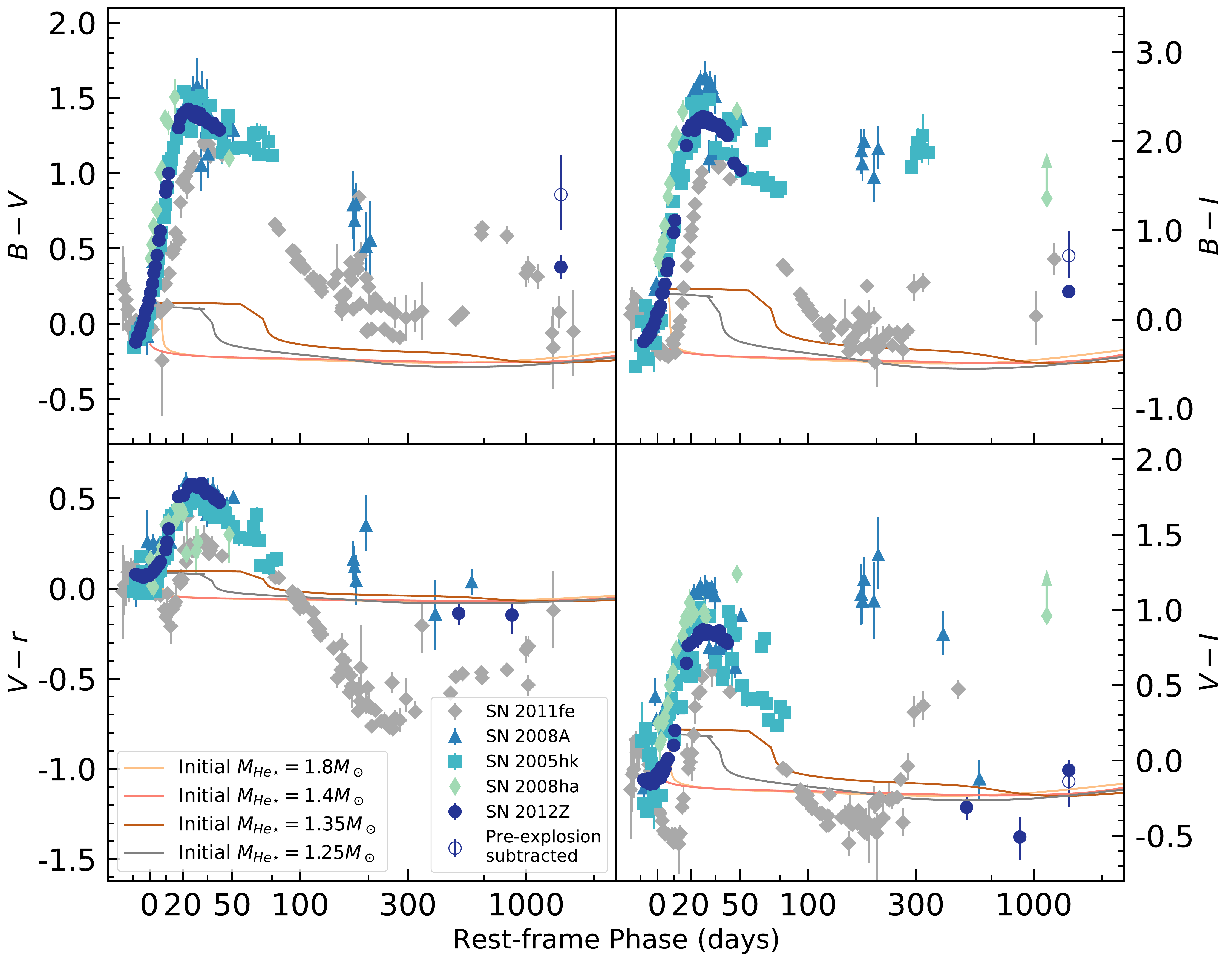}
\end{center}
\caption{Similar to Figure \ref{fig:colorcurves_remnant}, but comparing to models of the shock heated He star companion from \citep{Pan13}. The latest epochs of SN~2012Z are too red to be explained by the companion models so this suggests that the companion is not the dominant source of flux at these epochs.}
\label{fig:colorcurves_companion}
\end{figure*}

Another possibility to explain the late-time photometric behavior of SNe~Iax is the donor star. We know this star must be present in the system and \citet{Bulla20} find that including its contributions near maximum can better match the spectropolarimetry observations of SN~2005hk. \citet{Pan13} model the effects of the SN explosion on a He star companion. They find that the post-impact companion expands and goes through a luminous phase We compare our data to their models in Figure \ref{fig:colorcurves_companion}. We find that the companion models are too blue to explain our latest photometry, so the companion is likely not the dominant source of the flux at these epochs which is also supported by our estimates of the luminosity below.

If material was ejected from the binary system during its evolution to SN~2012Z, we would expect that CSM to eventually dominate the flux \citep{Gerardy04, Graham19}. CSM interaction models generally predict blue emission so if we assume the unsubtracted scenario, our latest color measurement is consistent with the CSM interpretation but more detailed modeling would be necessary to test if these models match in detail.

\subsection{Estimating the Optical Luminosity}
\begin{figure*}
\begin{center}
\includegraphics[width=\textwidth]{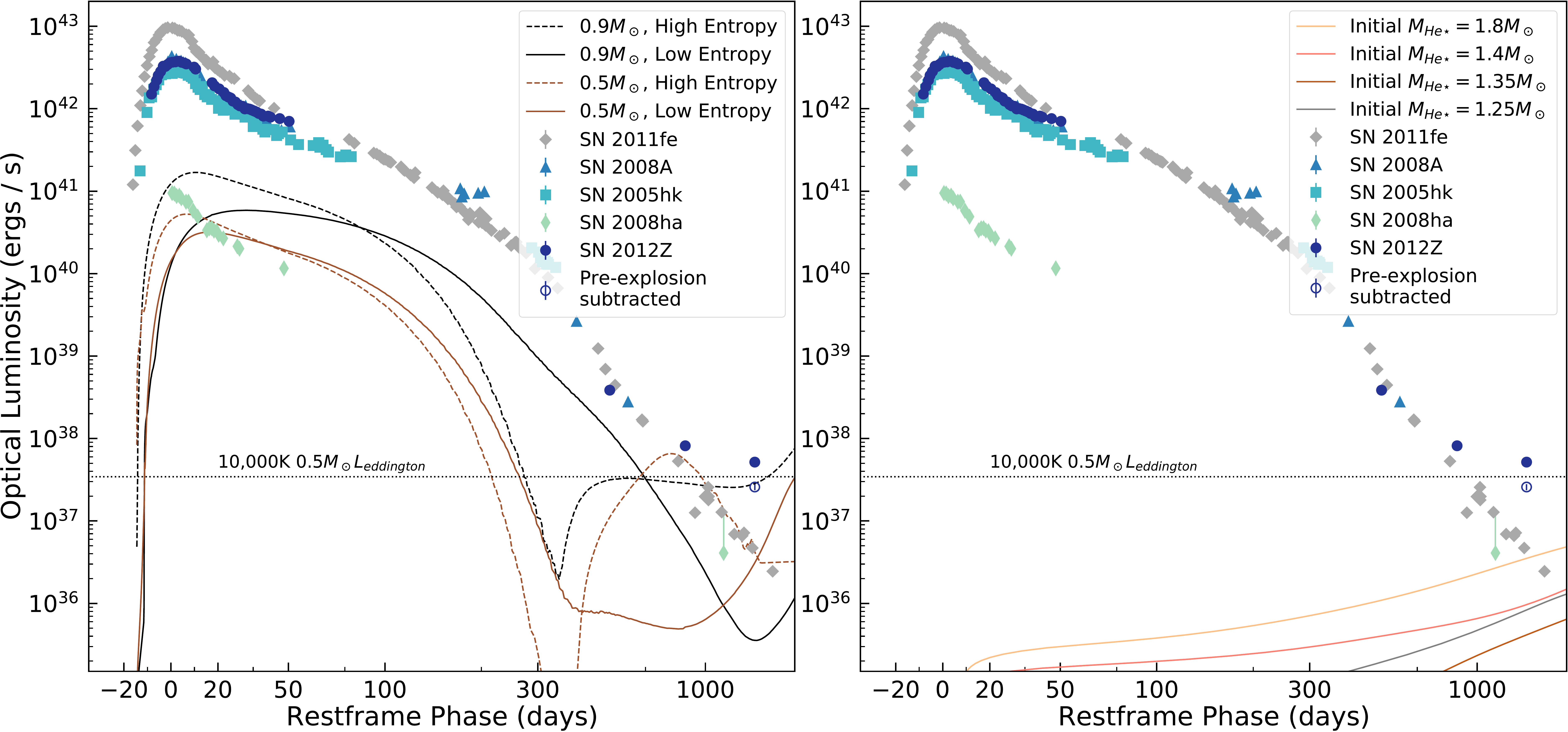}
\end{center}
\caption{\label{fig:bolometric} Optical luminosity of SN 2012Z (blue circles), SN 2005hk (light blue squares), SN 2008A (light blue triangles), SN~2008ha (green diamonds) and SN 2011fe (gray diamonds). SN 2012Z was brighter than SN 2005hk and SN 2008A near peak, but about one year after peak, SN 2008A and SN 2005hk were both brighter than SN 2012Z. We include both the pre-explosion subtracted (unfilled) and raw photometry (filled) to illustrate both scenarios. The left panel compares the the optical luminosity of the bound remnant models from \citet{Shen17} while the right compares to the heated companion star from \citet{Pan13}. SN 2011fe was nearly twice as bright as SN 2012Z at peak, but the light curves crossed at $\sim 850$ days. Later than 800 days, the decline of SN 2012Z has slowed considerably while SN 2011fe's has continued to decline at a roughly constant rate. The remnant models cool and become very red about a year past maximum where most of their flux drops out of the optical bands. At early times, neither the remnant nor the companion models produce enough flux to explain the data even to within a factor of 2. At the latest epoch, the remnant models reheat and produce flux that is comparable to what we see in our latest observation. The companion models are much fainter than our observations at the observed epochs. The companion star may be visible in later observations that will be presented by (Camacho-Neves et al.~in prep.).}
\end{figure*}

To estimate the ``optical'' luminosity, we use the SED models described above, integrating using the trapezoid rule. Similarly to \citet{McCully14_08A}, we integrate from 340--970 nm, where our data is best sampled. By only considering the optical range, we remove some of the systematics associated with extrapolating to get the full bolometric luminosity. We use the Monte Carlo samples to estimate the uncertainties in luminosity measurements.  The results are shown in Figure \ref{fig:bolometric}.

Near peak, SN 2012Z was brighter than SN 2005hk and SN 2008A. However, this reversed about 250 days past maximum: both SN 2008A and SN 2005hk were brighter than SN 2012Z at these epochs. All three SNe Iax (SNe 2012Z, 2008A, and 2005hk) have similar rise times, but SN 2012Z has slower decline post-peak. 

SN 2011fe was (unsurprisingly) brighter than all of the SNe Iax by about a factor of two at peak. However, at $\sim 850$ days past maximum the light curves of SN 2012Z and SN 2011fe crossed. SN 2012Z is approaching a constant luminosity while SN 2011fe continues to fade. At the latest epoch we have for SN 2012Z, nearly 1500 days after maximum, SN 2012Z is an order of magnitude brighter than SN 2011fe and is roughly at the Eddington Luminosity for a mass of 0.5 $M_\odot$.

The models from \citet{Shen17} are also shown on Figure \ref{fig:bolometric} in the left panel. Between 1 and 3 years after maximum, the models become so red that much of the flux has dropped out of the optical bands. It is not until our latest epochs that the models begin to heat and the flux moves back into the optical bands. The low entropy remnant models are too faint for our observations, but the high entropy models are similar to the final photometric observations for SN 2012Z.

An alternative explanation for the excess flux is that, instead of seeing a bound remnant, the flux is dominated by the shock-heated companion star. \cite{Pan13}, \cite{Shappee13}, and \cite{Liu13} predict that when the ejecta collide with the companion star, they will strip some mass and shock-heat the envelope. \citet{Pan13} calculated the properties of a surviving helium-star companion. They found that that the helium star would be hot ($\sim 30,000$ K) and bright ($\sim 10,000~L_\odot$). This corresponds to an absolute magnitude in $V$ of -5.2, within a magnitude of what we observe at our latest epoch. This is also consistent with the blue SED we observe (without subtracting the pre-explosion flux).

We compare our observations to the companion models from \citet{Pan13} in the right panel of Figure \ref{fig:bolometric}. We find that the models here are much fainter in the optical than we observe. The models do predict that the companion should become brighter at later epochs, so the companion may produce the dominant contribution of the flux in later observations perhaps even causing a rebrightening of the source.

\begin{figure}[!t]
    \centering
    \includegraphics[width=\columnwidth]{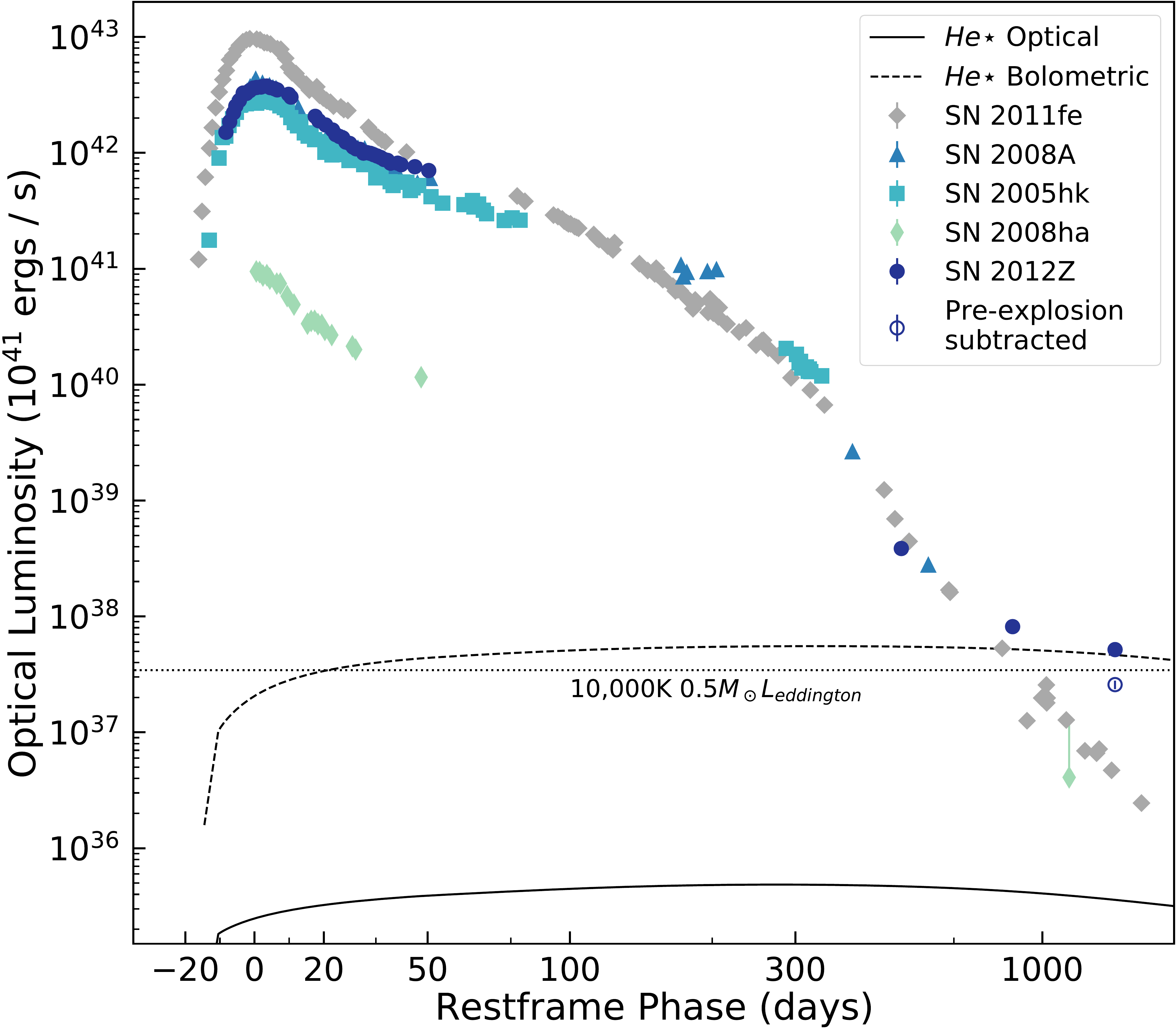}
    \caption{Optical luminosity of SNe Iax; Similar to Figure \ref{fig:bolometric}. The solid line shows the optical luminosity from the helium star companion from \citet{Zeng2020}. The optical luminosity is much too low to contribute our late-time observations. The dashed lines shows the bolometric luminosity from the same model and is quite similar to the flux we observe at $1450$ days past maximum. If that light was reprocessed into the optical by surrounding material, the companion could provide a significant fraction of the luminosity we observe at these late phases.}
    \label{fig:yaotian}
\end{figure}

In Figure \ref{fig:yaotian}, we compare our observations to the models from (\citealt{Zeng2020}; evolution of the post SN companion will be presented in Zeng et al.~in prep.). \citet{Zeng2020} use the 3D Smoothed Particle Hydrodynamics code Stellar GADGET \citep{Pakmor12}. Their models use the results from \citet{Liu13} as the initial state of the He star. \citet{Zeng2020} assume the \textit{N5def} model from \citet{Kromer13} who find that model best explains the observations of SN~2005hk. The \textit{N5def} is a weak deflgration of a Chandrasekhar-mass WD that ejects $0.372 M_\odot$ of material leaving a $1.03 M_\odot$ bound remnant. After the SN explosion, the evolution of the companion star is modelled using the 1-D MESA stellar-evolution code \citep{Paxton11, Paxton15, Paxton18, Paxton19}.

The solid line shows the optical flux which is too faint to contribute to our observations significantly. However, the bolometric luminosity from the models are comparable to our observed late-time flux. If there was material to redistribute the blue flux into the optical (e.~g.~ from a wind from a bound remnant), the shock heated companion could contribute to our late-time observations.

\subsection{Estimating Possible CSM Properties}
During the evolution of the binary leading up to SN~2012Z, material may have been escaped the system. When the ejecta collides with this material, a shock is produced creating another possible source of luminosity. The flux produced by the CSM interaction depends on the velocity of the shell and the mass accumulated by the ejecta. \citet{Jacobson-Galan19} find evidence for CSM interaction in some SNe Iax, so we estimate the order of magnitude properties of the CSM that would be necessary to explain the slow decline in our latest photometry. 

We consider the constant density CSM model from \citet{Gerardy04}. SN~2012Z has reached $\sim5\times 10^{37}$ ergs / s at the latest phases as shown in Figure \ref{fig:bolometric}. This is $\sim 6$ orders of magnitude fainter than what \citet{Gerardy04} considers for SN~2003du. If we assume that the luminosity scales linearly with density, this would imply a density of $1.66 \times 10^{-22}$ g / cm$^3$. At 1450 days of expansion at $8,000$ km / s, we would estimate a radius of $10^{17}$ cm. Integrating over the sphere, this would imply a total accumulated mass of $3.5\times 10^{-4} M_\odot$.

\citet{Smith16} write
\begin{equation}
    L = 2 \pi R^2 \rho v_{\rm{shell}}^3.
\end{equation}
Using the values above, this would imply a shell velocity of $1700$ km / s which is similar to the escape velocity of the progenitor system making this a plausible explanation for the late-time flux we see for SN~2012Z. We note that if this is the source of the excess late-time luminosity of SN~2012Z, then even this small amount of material cannot be present in the normal SN Ia, SN~2011fe as its light curve continues to decline following radioactive decay. More detailed modeling is necessary to distinguish whether the CSM interaction or a bound remnant dominates the luminosity at our latest observations.

\section{Discussion}\label{sec:discussion}

\subsection{Ejecta Mass}
There is significant evidence from our pre-explosion imaging of SN~2012Z \citep{McCully14_12Z} and from late-time spectra \citep{Foley16} that SNe Iax arise from single-degenerate, Chandrasekhar mass system. The ejecta mass is more uncertain. Here we consider models to guide our interpretation of our latest photometry.

Using a simple Arnett's scaling relation ($M_{\rm{ej}} \propto v t_{\rm{rise}}^2$) that is only based on the ejecta velocity and the rise time \citep{Arnett82}, we find that SN~2012Z has about a third of the ejecta mass of SN~2011fe which would give $\approx0.5~M_\odot$. This would imply a $0.9~M_\odot$ remnant if the WD exploded at the Chandrasekhar mass. 

However, while the light curve of SN~2012Z has a fast rise time, it is also broad after maximum, ($\Delta m _{15,bol} = 0.6 $ mag, $\Delta m_{40,bol} = 1.6 $ mag) that would for a normal SN Ia suggest a super-Chandrasekhar ejecta mass, $M_{\rm{ej}}$, of 1.7 -- 2.0 $M_\odot$ \citep{Scalzo19}. This is consistent with the estimates of \citet{Stritzinger15} for SN~2012Z who use a similar technique. However, this model does not take into account a bound remnant and may differ for SNe Iax given their density profile.

To explore this, we reexamine the hierarchical Bayesian model of \citet{Scalzo14_1, Scalzo14_2, Scalzo19}. The model includes a semi-analytic prediction of radioactive energy deposition within a spherically symmetric, homologously expanding SN Ia ejecta with a user-defined radial density profile and $^{56}$Ni distribution \citep{Jeffery99}. The ejecta expansion velocity is determined by energy balance, subtracting the binding energy of the white dwarf \citep{Yoon05} from the nuclear energy released in conversion of carbon and oxygen to the final ejecta composition \citep{Maeda09}. These models are similar to those from Arnett, but use model driven priors and marginalize over unknown parameters values rather than assuming a simple scaling as we did above. The Arnett scaling relation for estimating the ejecta mass is based on photospheric-phase behavior, relying on measurements of the rise time and the velocity of the Si II absorption feature at maximum light. In contrast, the Scalzo parametric model relies on bolometric measurements made at later phases when the ejecta have become optically thin to $^{56}$Co gamma rays (from $+40$ -- $+100$ days).

The potential for a bound remnant complicates this interpretation and so the semi-analytic explosion model must be updated to account for this in some way. We incorporate into the framework a schematic treatment of the mass of a bound remnant. The remnant mass contributes to the progenitor binding energy but is not part of the ejecta mass available for scattering of $^{56}$Co gamma rays from the ejecta. The binding energy of the remnant cannot be turned into the kinetic energy of the ejecta so the remnant thus acts as a drag on the expanding ejecta, lowering the expansion velocity. Since numerical simulations indicate a puffed-up remnant relative to a normal white dwarf \citep{Kromer12, Kromer13, Zhang19}, we approximate the final remnant binding energy as being negligible for our estimate. The effect of the remnant on this simplified explosion is thus to lower the expansion velocity of the ejecta, producing a flatter bolometric light curve past peak. We produce simulations using both ejecta radial density profiles used in \citet{Scalzo14_1}: a profile that is exponential in velocity and a power-law density profile tuned to match 3-D simulations of SNe Ia \citep{Kromer10, Pakmor12, Seitenzahl13}. 

The ratio of the observed bolometric luminosity to the modeled luminosity from radioactive decay ($\alpha$) is a nuisance parameter in our analysis, but is expected to be close to 1 \citep{Branch92, Stritzinger06, Howell06, Howell09}. In the semi-analytic models of \citet{Arnett82}, $\alpha = 1$ identically.  We consider two priors for alpha, as in \citet{Scalzo14_1}:  a Gaussian prior $\alpha = 1.00 \pm 0.01$, corresponding to the ideal Arnett case, and a Gaussian prior $\alpha = 1.2 \pm 0.2$, corresponding to the range seen in the numerical models of \citep{Hoeflich96}.

The presence of an unknown amount of material with gravitational influence but no effect on gamma-ray opacities requires additional priors to constrain meaningfully. In contrast to previous analyses of normal SNe Ia \citep{Scalzo14_1, Scalzo19}, we introduce a Gaussian prior on the bulk kinetic energy velocity based on the Si II 6355 line velocity at maximum light with a standard deviation of $500 $~km~s$^{-1}$.  We adopt the following silicon velocity priors: SN~2011fe: 10,400~km~/~s \citep{Zhang16}; SN~2012Z: 7,500~km~/~s \citep{Stritzinger15}; SN~2008A: 7,000~km~/~s \citep{McCully14_08A}; SN~2008ha: 3,700~km~/~s \citep{Foley10_08ha}; and SN~2005hk: 5,500~km~/~s \citep{Phillips07}.

\begin{figure*}
\begin{center}
    \includegraphics[width=\textwidth]{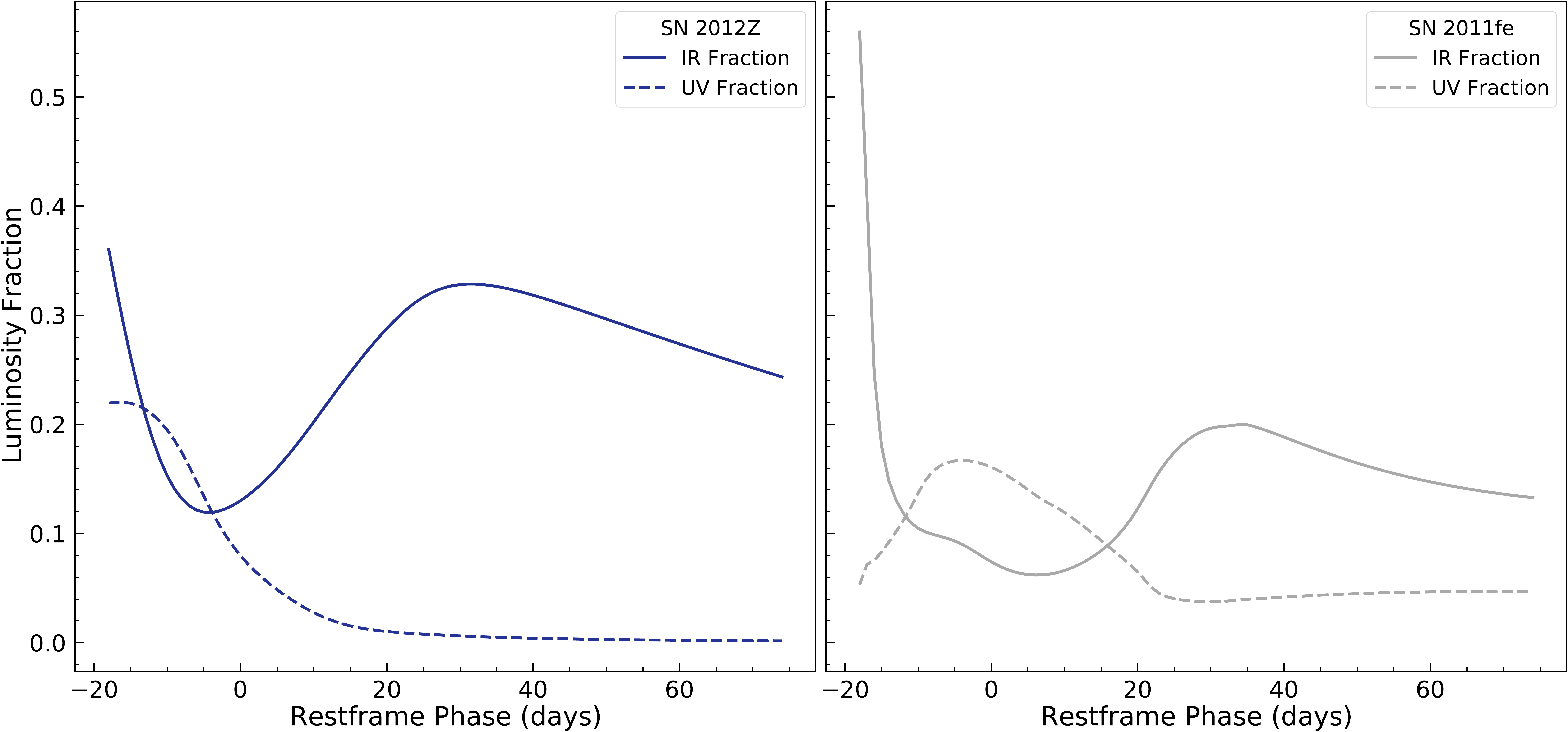}
    \caption{Bolometric luminosity contribution from the UV and IR for SN~2012Z and SN~2011fe. We fit a phenomenological light curve model from \citet{Bazin2009} to the light curve from each filter to interpolate the UV/IR contributions onto a uniform grid. Linearly interpolating in wavelength, we integrate from 160nm to 340nm for the UV, 340nm to 970m for the optical, and 970nm to 2.4$\mu$m for the IR. The UV and IR contributions we derive for SN~2011fe match those presented in \citet{Pereira13} to a few percent. When applying these corrections to the SNe Iax presented here, we a pre- and post-max stretch fit using the integrated optical light curve of SN~2012Z as the template.}
    \label{fig:uvir}
\end{center}
\end{figure*}

These models require the bolometric light curve rather than the integrated optical that we presented above. SN 2012Z has both UV and IR observations enabling a bolometric correction, but several of the other SNe Iax presented here do not. To account for this, we use IR observations from \citet{Yamanaka15} and Swift observations from \citet{Brown14} to estimate the bolometric correction for SN 2012Z (using the same method to convert from photometry to luminosity as above). To interpolate our observations onto a uniform grid, we fit the light curve in each band to a series of light curve functions from \citet{Bazin2009}. Using the light-curve fits, we integrate over wavelengths to get the luminosity using the pivot wavelengths as the filter centers. We then fit a pre-peak and post-peak stretch to the SN Iax integrated optical light curves and apply a stretched bolometric correction accordingly. We implement the same procedure using data from \citet{Matheson12} in the IR and Swift observations from \citet{Brown14} for the normal Ia SN~2011fe. Our estimated UV and IR contributions are shown in Figure \ref{fig:uvir}.

\begin{figure*}
\begin{center}
    \includegraphics[width=\textwidth]{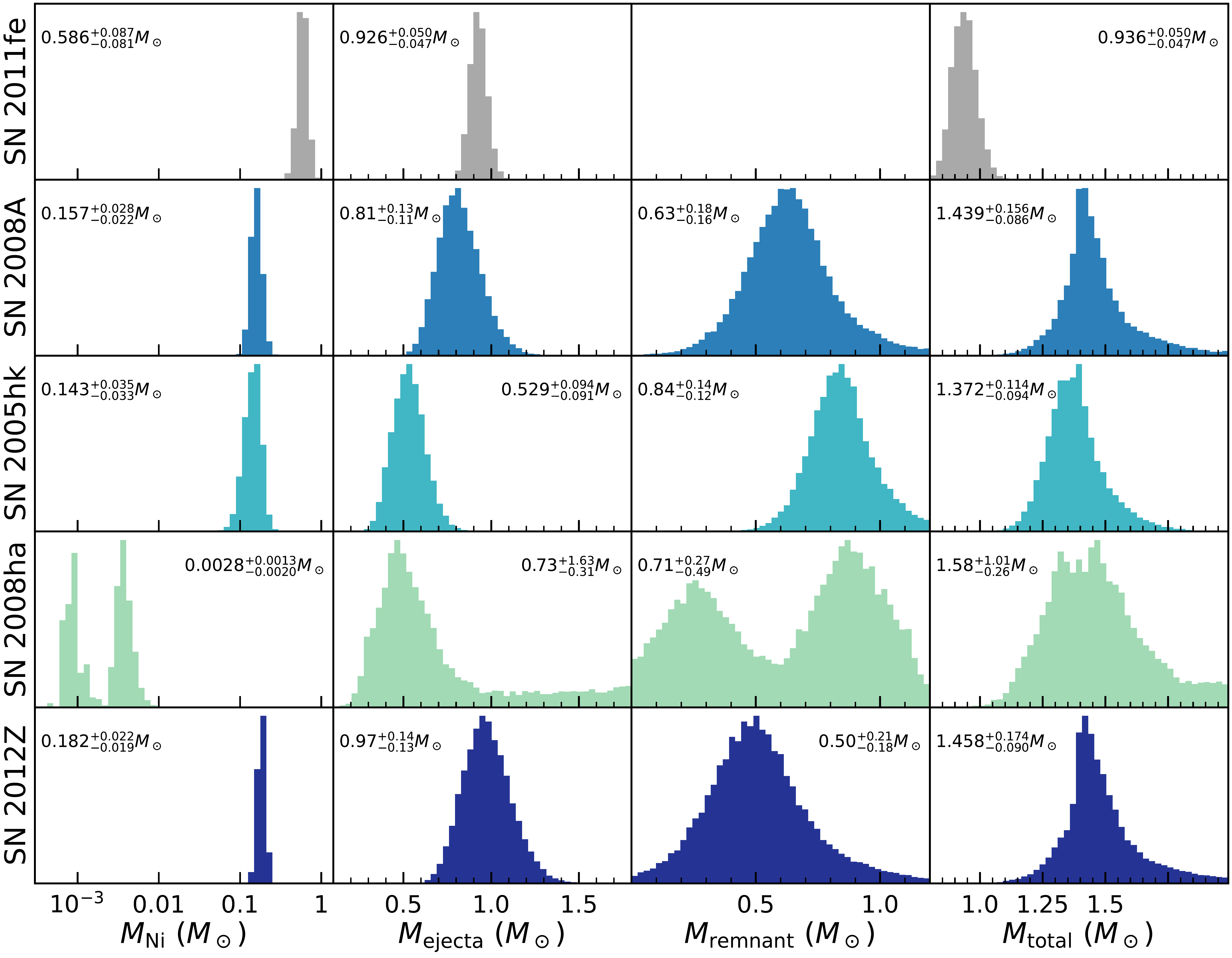}
    \label{fig:scalzo}
    \caption{Marginalized probability density functions (PDFs) from our hierarchical Bayesian models for the exponential density profile using the early light curves of the SNe presented throughout this work. Our models favor a sub-Chandra explosion for SN~2011fe. We explicitly assume a prior of zero remnant mass for SN~2011fe as it is a normal SN Ia. The nickel mass estimate is comparable to previous works, but ejecta mass is slightly less than that found by \citet{Scalzo14_1} who also argued SN~2011fe was sub-Chandra. Here we include a prior on the scale velocity derived from the peak silicon velocity whereas \citet{Scalzo14_1} only used light curve information. These differences imply a systematic uncertainty in our ejecta mass estimates of $\sim 0.1 M_\odot$. The models shown here are all peaked near the Chandrasekhar mass for total mass of the system. Models that have a power-law density distribution are systematically higher pushing the best fit models into the super-Chandra regime. The PDF for SN~2012Z peaks at $0.5 M_\odot$, but given the systematics, the models are consistent with but do not require a non-zero remnant mass for SN~2012Z.The models of SN~2005hk and SN~2008ha peak at a much higher remnant mass and favor a non-zero remnant mass even taking into acount systematics. Our models show a correlation between ejecta mass and nickel mass for SNe Iax but more detailed modeling is necessary to confirm this result.}
\end{center}
\end{figure*}

The results of the exponential density profile models are shown Figure \ref{fig:scalzo}. The results for SN~2011fe are shown in the top row. We have explicitly included a prior that SN~2011fe did not leave behind a remnant. The nickel mass we obtain is consistent with previous works (e.g. \citealt{Pereira13}). The ejecta mass and therefore total mass are both sub-Chandrasekhar. This is qualitatively consistent with \citet{Scalzo14_1} who find that SN~2011fe favored a sub-Chandra explosion at 1.08~$M_\odot$ for comparable models. More discussion of the reanalysis of SN~2011fe is in an appendix. The models here favor a lower ejecta mass of $0.95~M_\odot$. Here we are showing the exponential density distributions; the power-law density models favor a $0.1~M_\odot$ higher ejecta mass implying a $\sim 0.1~M_\odot$ systematic uncertainty in our ejecta mass estimates which must be accounted for in our interpretation of the models. Our estimates of the ejecta mass of SN~2011fe are consistent with those from the models of \citet{Shen21} within systematics.

We find that for these models of the SNe Iax, all of the total mass estimates are centered around the Chandrasekhar mass. There were no explicit constraints on the models to enforce this, so it is noteworthy that all of our models peak at this value. The power-law density models have total masses that are consistent with the Chandrasekhar mass but favor higher, super-Chandra total mass estimates similar to the findings of \citet{Stritzinger15} for SN~2012Z. Chandrasekhar WD masses are consistent with \citet{Foley16} who find that SNe Iax produce a significant amount of stable Ni which require high density burning. The material would need to burn near the center of the WD but then also escape as part of the ejecta. For example, \citet{Blondin2021} and \citet{Leung2020} show that deflagrations of Chandrasekhar mass WDs can produce a significant amount of stable Ni in the ejecta. In \citet{Leung2020}, 2-D simulations show that their models can produce both a remnant and stable Ni. Synthetic observables of their models are necessary to test if they can reproduce the layered spectral evolution seen in \citet{Stritzinger15} to see if they are viable models for SNe Iax.

The exponential density profile models shown have probability distributions that peak with SN~2012Z with the smallest remnant mass of $0.5^{+0.21}_{-0.18}~M_\odot$ and the highest remnant masses for SN~2005hk. The low velocities of SN~2005hk favor high remnant masses in our models. Given the systematics of at least a $0.1 M_\odot$, our models do not require a bound remnant but are consistent with one for SN~2012Z. However, the models for both power-law and exponential density profiles favor a non-zero remnant mass for SN~2005hk at higher than $95\%$ confidence.

Our models are not able to fit the light curve from SN~2008ha adequately to draw conclusions. SN~2008ha is the most extreme object in our sample and has the least data coverage because it was so faint. The models that converged (and are shown in Figure \ref{fig:scalzo}) require a large value of $\alpha$, implying less efficient conversion of radioactive decay to SN luminosity. The peak of the total mass is consistent with the Chandrasekhar mass, but the other parameters are multi-modal and have broad tails of their distributions. In the high-mass, low-56Ni tail of parameter samples for SN~2008ha, the inferred timescale for passing into the optically thin regime for 56Co gamma rays far exceeds the range of available data, breaking the main assumption used to derive our light curve model.  We therefore do not consider our model reliable for SN 2008ha, and further constraints or different methods are needed to draw any significant conclusions.

Our ejecta masses shown here are more similar to the lower ejecta masses predicted by only using the rise of the light curves of SNe Iax. We find a correlation for the four objects modeled here between the ejecta mass and the radioactive nickel mass. This is in line with the naive expectation that a more powerful explosion will produce more ejecta and produce more radioactive nickel. If the rise times are better indicators of the ejecta mass, this could explain correlation between rise time and luminosity found by \citet{Magee16}.

Further theoretical/explosion simulations are necessary to confirm these results but are beyond the scope of this work.

\subsection{Flux Contribution of the Remnant}
At this point, we have at least three possible sources for the brightness at our latest observation of SN~2012Z: Long half-life radioactive decay, CSM interaction, and a bound remnant. Here we further consider other signatures that could differentiate a bound remnant from the other models. If the flux of the remnant was substantial at phases $>60d$, it would make the lightcurve appear broader than it actually systematically biasing our ejecta model above. To estimate the contribution of a remnant to the SN photometry, we use SN~2011fe as a template light curve. We shift, scale, and stretch the optical light curve of SN~2011fe to match the SNe Iax at peak, and then subtract resulting lightcurve as an estimate of the flux from the SN ejecta. 

\begin{figure}
\begin{center}
\includegraphics[width=\columnwidth]{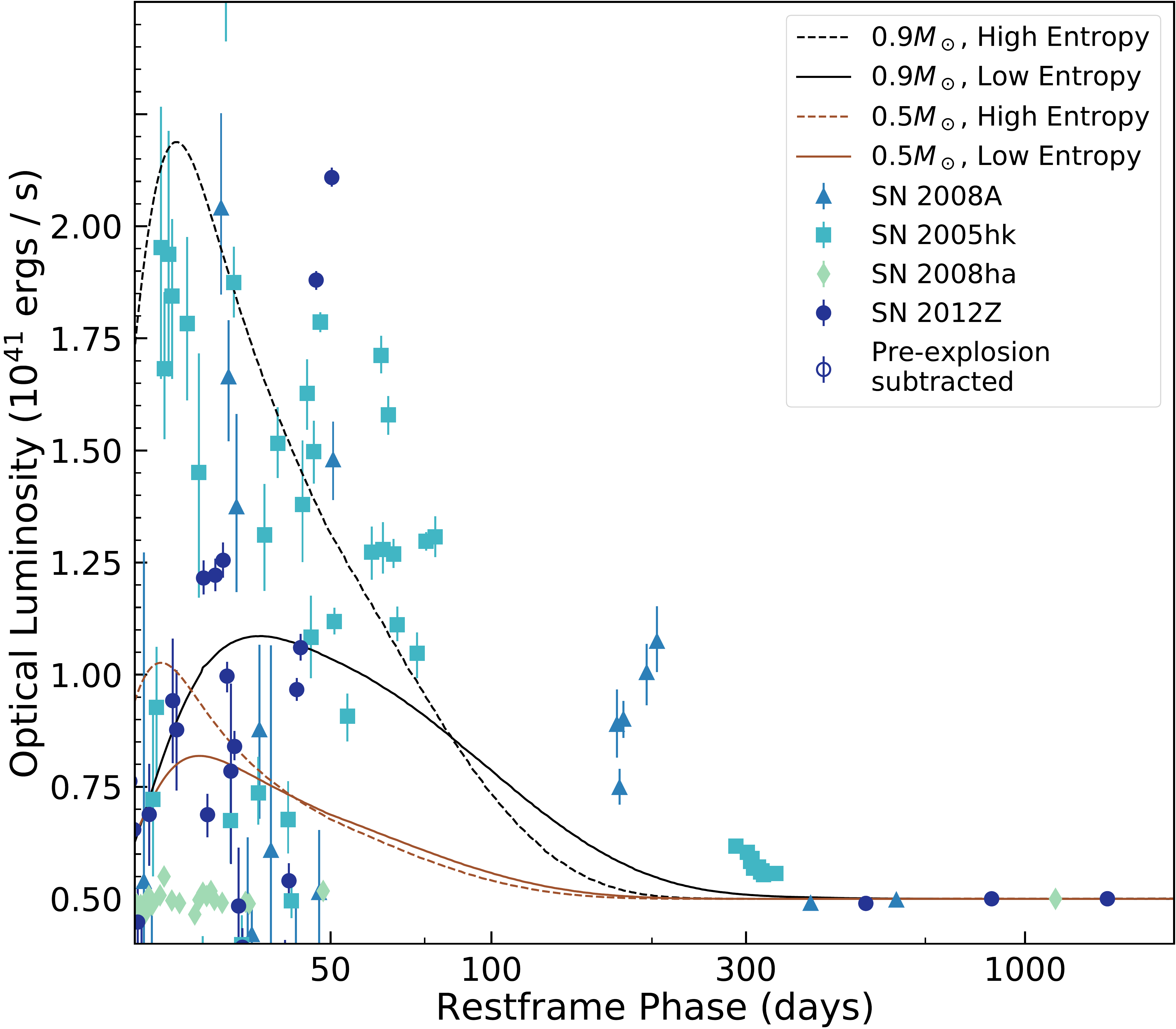}
\end{center}
\caption{\label{fig:bolometric_sub} Similar to Figure \ref{fig:bolometric}, but we have subtracted off a scaled, stretched light curve of SN~2011fe to estimate the contribution from the SN ejecta and from any possible remnant. Prior to 50 days after maximum, the data is noisy and non-constraining as the data taken near peak were not as deep as observations taken later. Data is sparse for most of the SNe at intermediate epochs ($>50$ days), but the data from SN 2005hk qualitatively follow the predictions of the bound remnant models from \citet{Shen17}. At these epochs, the companion star models from \citet{Pan13} are much fainter than the data so they are not included here.}
\end{figure}

Our results are shown in Figure \ref{fig:bolometric_sub}. Before 50 days after peak, the error bars are large because the data near peak was not as deep as observations taken later. Therefore, we do not draw any conclusions from this data. At later epochs, there is an excess for both SN~2005hk and SN~2012Z. SN~2005hk has by far the best sampling at these intermediate phases and qualitatively follows the prediction of remnant models. This ``ejecta + remnant'' model is the only one we have tested that can explain the entire evolution of the bolometric light curve. More realistic models of the light curve from the ejecta that explore Ni distribution and density profile effects are necessary to confirm the robustness of this explanation. 

We conclude that models that rely on data at phases later than 50 days can be potentially contaminated by a remnant and will bring down the estimates of ejecta mass estimates from these models.

One model that can produce a remnant is the pure deflagration of a WD \citep[e.~g.~][]{Kromer13}. However, these models are not without their issues. While the peak of the light curves from \citet{Kromer13} and \citet{Kromer15} match SN~2005hk well, just 30 days past peak they under-predict the flux. Based on our analysis shown in Figure \ref{fig:bolometric_sub}, one possibility is that the lightcurve of SN~2005hk is already starting to be contaminated by the remnant at these phases. There are other open questions about the pure deflagration models like if they can produce the layered ejecta seen in SN~2012Z \citep{Stritzinger14, Barna18}, but this is beyond the scope of this work. 

A general feature of the remnant models from \citet{Shen17} is that from about 300 days until $1000--2000$ days the remnant is very red with most of its flux emitted in the IR. The higher mass remnants (lower ejecta masses assuming Chandrasekhar mass) tend to say red until later epochs than lower mass remnants. It is therefore possible that the source that we found to be coincident with SN~2008ha in \citet{Foley14} is the remnant still in the red phase before the remnant reheats, though detailed modeling would be necessary to confirm this hypothesis.

The light curve of SN~2012Z (as well as of other SNe Iax) is roughly an order of magnitude brighter than the remnant models a few hundred days past maximum. This is a significant issue for the remnant models given that we would expect the remnant brightness to be comparable to the ejecta at $\sim 300$ days past maximum if the spectra are a composite of forbidden emission features from the ejecta and P-Cygni lines from the remnant wind. This may be less of a problem for SN~2012Z if the ejecta is simply brighter than that of SN~2005hk, but the velocities of SN~2012Z are much higher than for SN~2005hk \citep{Foley16} making it difficult to distinguish if the Fe P-Cygni features identified in SN~2005hk \citep{McCully14_08A} are faint or are just blended. It is possible the continuum estimations in the existing remnant models are too simplistic and the emission lines are stronger than expected but detailed radiative transfer would be needed to test this idea and is beyond the scope of this work.

We encourage others to use our observations as constraints on their simulations to better understand the physical mechanisms that produce SNe Iax.

\section{Conclusions}\label{sec:conclusions}
We have presented extremely late-time photometry of SN 2012Z, one of the brightest SNe Iax, taken with \textit{HST} from 502 days to 1425 days past maximum light. We find that the light curve decline has slowed unlike the normal type~Ia SN~2011fe, which continues to decline at a higher rate. While SN 2011fe was a factor of 2 brighter than SN 2012Z at peak, SN 2012Z is now brighter than SN 2011fe by a similar factor at 1425 days past maximum light. 

Empirically, the late-time light curve of SN~2012Z is well fit by an exponential decay model in magnitude units. The asymptotic limits of these models are within $2-\sigma$ of our latest photometry though the future observations may deviate from these curves as different power sources become dominant. Surprisingly, these asymptotic fluxes are still brighter than the pre-explosion detection in all filters by at least a factor of two. At all epochs, the SED is blue and consistent with being thermal. 

We expect that this flux is a composite of several sources: the shock-heated companion, a bound remnant that could drive a wind, light from the supernova ejecta due to radioactive decay, and CSM if it is present. Our measurements were brighter than the predicted radioactive decay models for $^{56}$Co and $^{57}$Co so if the light curve is dominated by radioactive decay, it must be from longer lived species like $^{55}$Fe. The models of the shock-heated companion that we have compared to here suggest that the contribution from the companion star is too blue to be observed in the optical but could be reprocessed into the optical and the flux from the companion star could increase at later phases. The models from \citet{Shen17} of a bound remnant can produce similar luminosities to what we observe at our latest epochs, but interaction with CSM could also play an important role at these epochs.

Our theoretical models of the early light curves all peak at a Chandrasekhar mass for SNe Iax and are consistent with a bound remnant. Models that produce synthetic observables to compare to the observations presented here are essential to understand the physical mechanism that produces SNe Iax. 

Our models peak at 0.94 $M_\odot$ for the normal Ia SN~2011fe favoring a sub-Chandra WD.

Models suggest that we have not yet seen the contribution from the shock-heated companion star, but data taken a few years after that presented here could show a rebrightening as the companion expands. We will test this hypothesis using \textit{HST} observations at $\sim 2500$ days which will be presented in Camacho-Neves et al.~in prep. These observations will help complete the picture of the full life cycle of SNe Iax.

\begin{appendix}
As part of this work, we reanalyzed the light curve of the normal Ia SN~2011fe to make a uniform comparison between SNe Iax and normal SNe Ia. The ejecta masses derived here are lower than those found in \citet{Scalzo14_1}. We discuss the differences here.

The differences in priors between our analysis and \citet{Scalzo14_1} can account for some of the differences we see in our inferences about SN 2011fe. \citet{Scalzo14_1} did not use a prior on the scaling velocity based on the observed Si II line velocity near maximum light, whereas in this work it is necessary in order to provide useful constraints on possible remnant masses.  \citet{Scalzo14_1} also found that looser priors on the assumed mass fraction of unburned carbon/oxygen in the outer layers tended to reduce the inferred ejecta mass; their fiducial analysis uses M$_{\rm{CO}} < 0.05$ M$_{\odot}$, while this paper uses M$_{\rm{CO}}$ = $0.2 \pm 0.1$ M$_\odot$.

There are also differences between the bolometric light curve presented here and the one from \citet{Scalzo14_1} based on how corrections were made for flux outside the optical. \citet{Scalzo14_1} used a NIR correction template, parameterized by decline rate, based on the first NIR data release of the Carnegie Supernova Project \citep{Folatelli2010, Stritzinger2011}: in this respect SN 2011fe was analyzed in the same way as the other SNfactory SNe Ia which lacked extensive NIR data. The NIR template correction used in \citet{Scalzo14_1} at 40 days past maximum light was between 25\%-30\% for SNe Ia with SALT2 $x_1 \sim 0$, while our measured NIR flux fraction for SN~2011fe at day +40 is 19\%. \citet{Scalzo14_1} also did not include a template correction for unobserved UV flux; instead the UV flux fraction at maximum light was treated as a nuisance parameter in the statistical model with a uniform prior between zero and 10\%, and marginalized over.  This work finds a UV flux fraction of 16\% at maximum light. Both of these differences in the light curve tend to decrease the inferred mass due to the additional NIR flux on the post-photospheric luminosity, and the higher inferred $^{56}$Ni mass driven by the new UV correction which requires a lower gamma-ray optical depth to reproduce the observed luminosity at day +40.

When re-analyzing the \citet{Scalzo14_1} bolometric light curve of SN 2011fe based on the priors and Swift-based UV flux fraction from this paper, we found a mean decrease of 0.16 $M_\odot$ across all assumed distributions of the ratio of bolometric to radioactive decay luminosities ($\alpha$) and radial density profiles. That decreased value is now consistent with what is found in this work.
\end{appendix}

\begin{acknowledgements}
We thank Wen-fai Fong and Max Stritzinger for help with this original proposal for this data. We thank J.~Craig Wheeler and Stefano Valenti for enlightening conversations about the ejecta mass calculations. We thank the referee for their comments.

GH, CM, and DAH are supported by NSF grant AST-1313484. This research at Rutgers University (SWJ, YC-N) was supported by NASA HST programs HST-GO-12913, HST-GO-13360, and HST-GO-13757, as well as NSF grant AST-1615455.
RJF\ is supported in part by NSF grant AST-1518052, the Gordon \& Betty Moore Foundation, the Heising-Simons Foundation, and by a fellowship from the David and Lucile Packard Foundation. ZWL is supported by the National Natural Science Foundation of China (NSFC, No.11873016), the Chinese Academy of Sciences, and Yunnan Province (Nos. 202001AW070007).

These results are based on observations with the NASA/ESA Hubble Space Telescope and obtained from data hosted at the Space Telescope Science Institute, which is operated by the Association of Universities for Research in Astronomy, Incorporated, under NASA contract NAS5-26555. Support for Program numbers HST-GO-12913, HST-GO-13360, and HST-GO-13757 was provided through a grant from the STScI under NASA contract NAS5-26555.

This research has made use of the NASA/IPAC Extragalactic Database (NED), which is operated by the Jet Propulsion Laboratory, California Institute of Technology, under contract with the National Aeronautics and Space Administration.
\end{acknowledgements}

\facilities{HST (ACS, WFC3), NED}
\software{\url{https://github.com/cmccully/snhst} \citep{snhst}, Astropy \citep{astropy:2013, astropy:2018}, Dolphot \citep{Dolphot}, IRAF: IRAF is distributed by the National Optical Astronomy Observatory, which is operated by the Association of Universities for Research in Astronomy (AURA) under a cooperative agreement with the National Science Foundation., astro-scrappy \citealt{Astroscrappy}, numpy \citep{numpy}, scipy \citep{scipy}, drizzlepac \citep{Drizzlepac}, emcee \citep{emcee}, matplotlib \citep{matplotlib}}
\bibliographystyle{aasjournal}
\bibliography{late_sn2012z}
\end{document}